# Non-local Chemistry Driven by Cation-Anion Size Disparity in Helium Inserted Compounds under High Pressure

*Zhen Liu,[1,2] Stefano Racioppi,[3] Katerina P. Hilleke,[3] Abhiyan Pandit,[4] Shuran Ma,[1] Andreas Hermann,[5] Dadong Yan,[6] Eva Zurek,[3] Mao-sheng Miao[4*]*

1. *School of Physics and Astronomy, Beijing Normal University, Beijing 100875, China*
2. *Key Laboratory of Multiscale Spin Physics (Ministry of Education), Beijing Normal University, Beijing 100875, China*
3. *Department of Chemistry, State University of New York at Buffalo, Buffalo, New York 14260-3000, United States.*
4. *Department of Chemistry and Biochemistry, California State University Northridge, California 91330-8262, USA*
5. *Centre for Science at Extreme Conditions and SUPA, School of Physics and Astronomy, The University of Edinburgh, Edinburgh EH9 3FD, United Kingdom*
6. *School of Physics, Zhejiang University, Hangzhou 310058, China*

**Abstract**

Opposing the theory that Helium (He) cannot be inserted into AB-type ionic compounds due to the Madelung energy increase, our crystal structure search and first-principles calculations found that He can form stable compounds with sodium halides (NaX, X=Cl, Br, I) under high-pressure. These reactions are driven by the non-local chemistry arising from the cation-anion size disparity, distinctly different from the He insertion reaction with $A_2B$-type compounds. The large size differences between $Na^+$ and $X^-$ enable structures that can effectively host He insertions through volume and inter-atomic distance disproportionation. Furthermore, the insertion of He atoms can significantly relieve the elevated Madelung energy that builds up in NaX under high pressure. This energy increase arises from structural transitions driven by cation-anion size disparity, which are necessary for reducing volume under pressure. The insertion of He allows the reduction of the total volume under high pressure without increasing the Madelung energy. Our predicted compounds and stability analysis reveal a new example of He reactivity governed not by local chemical bond formation, but by long-range electrostatic interactions.

## Introduction

Helium (He) is the lightest noble gas (NG) and the most chemically inert element. Electrons are hard to transfer from or to He because the $1s^2$ closed shell configuration is associated with an extremely high ionization potential and low electron affinity. Thus, the chemistry of He differs substantially from that of the heavier NG elements such as Kr and Xe. The latter NGs exhibit very rich chemistry in the solid state, as exemplified by a large number of synthesized compounds.[1–15] A series of recent experimental and computational studies showed that pressure could greatly enrich the chemistry of heavy NG.[16–19] As a matter of fact, almost all types of chemical bonds can be found in various high-pressure NG compounds, including cationic and anionic NGs, NG-NG covalent bonds, and halogen bonds, NG analogue of the H-bonds).[20,21]

In contrast to heavy NGs, the high-pressure methods seem not very effective in making He compounds,[22–24] and when they do, the chemistry driving compound formation seems very different. For example, it has been theoretically predicted[25] and then experimentally shown[26,27] that He can be adsorbed by porous compounds under low pressures (below 1 GPa), and this effect was attributed to an increase in the entropy of the helium-stuffed systems resulting from helium's high compressibility. Among all the elements, He was found to only form thermodynamically stable compound with Na under pressures above 113 GPa.[24] The driving force for the reaction stems from sodium's propensity to become an electride under pressure, making it analogous to an ionic compound with a non-1:1 ratio of cations ($Na^+$) and anions ($E^{2-}$, which refers to a pair of electrons localized at an interstitial region). Later work demonstrated that the insertion of He into non-1:1 ionic compounds, such as $Na_2O$ and $Na_2S$, could lower the electrostatic energy increased under pressure.[28] This mechanism was extended and can be used to explain and predict many He insertion reactions with various compounds under pressure, including ionic compounds and polar molecules.[29–31] Based on this theory, He insertion into AB-type ionic compounds with 1:1 cation and anion ratio is not favored because it will increase the Madelung energy.[32]

It was, therefore, quite surprising when our calculation results showed that He reacts and forms stable compounds with Na halides. Our large-scale crystal structure search that explored He-NaX systems with various numbers of atoms in the primitive cell found stable NaXHe (X = Cl, Br and I) compounds with 1:1:1 molar ratio. Similar to previous studies,[32–34] electronic structure calculations showed that the He atoms inserted into sodium halides at high pressure do not donate nor accept distinguishable amount of electron with neighboring atoms. In-depth electronic structure and bonding analyses reveal that the insertions of He into NaX are driven by the large size disparity between $Na^+$ and $X^-$ ions in these compounds. On the one hand, when the size differences between the cations and anions comprising NaX is large, the Na+ ions are forced close to one another under pressure increasing the Madelung energy, because of the need to reduce the volume under pressure. The insertion of He atoms between the too-close cations can then relieve the electrostatic repulsion. On the other hand, the large size disparity enables NaXHe to develop highly symmetric structures with significantly different interstitial volumes. This interstitial volume disproportionation allows He to occupy the larger interstitial sites in NaX, producing a NaXHe structure with minimum total volume that is favored under high pressure. Our work reveals a new type of insertion of He into ionic compounds, caused by a very different chemical mechanism.

## Results

**Stability and structures of He insertion into sodium halides** The feasibility of the reaction between He and NaX, (X = F, Cl, Br and I) in a 1:1 ratio was studied by calculating the enthalpy of the reaction resulting in the insertion of helium into these four sodium halides in the pressure range of 0 - 300 GPa, with 50 GPa intervals. The reaction enthalpy per formula unit ($\Delta H$) is defined as the enthalpy of NaXHe as compared to the sum of the enthalpies of the sodium halide and elemental He as:

$$\Delta H = H[\mathrm{NaXHe}] - \{H[\mathrm{NaX}] + H[\mathrm{He}]\}. \tag{1}$$

Experimentally, the face-centered-cubic (FCC) to hexagonal-close-packed (HCP) phase transition in elemental He occurs in the pressure ($P$), temperature ($T$) regime where $P$ is between ~1 and 10 GPa, and $T$ is between 15 and 285 K.[35] Our calculations employed elemental He in the FCC structure at ambient pressure and in the HCP structure at high pressure. We found the enthalpy differences between these two structures to be negligible, and the choice of the reference structure did not change the trends and conclusions discussed here. The reference phases chosen for NaX are more complex and will be discussed in Section 3.2. All calculations are performed at zero temperature, and assuming static nuclei. The temperature effects will be discussed in a future work. NaFHe was not found to be enthalpically stable up to 300 GPa, whereas NaClHe, NaBrHe and NaIHe were found to be enthalpically preferred above 62, 38, and 44 GPa, respectively (Figure 1). While the reaction enthalpies decrease monotonically for NaClHe and NaBrHe in the studied pressure range, a minimum in the reaction enthalpy is observed for NaIHe, roughly -0.13 eV per f.u., at 200 GPa, further increasing to about -0.05 eV per f.u. at 300 GPa.

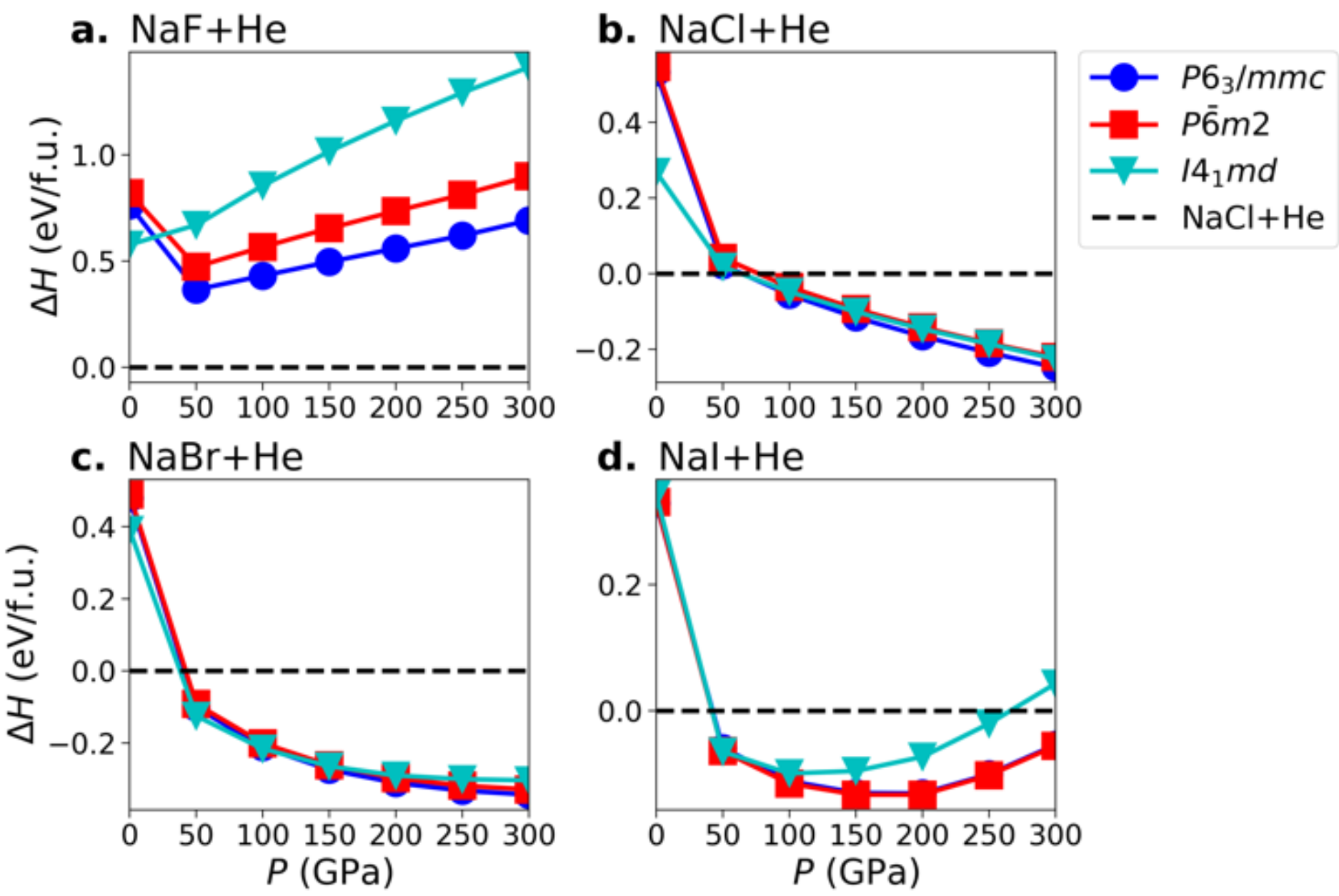

**Figure 1. The reaction enthalpy of He insertion into various sodium halides**: **a.** NaF, **b.** NaCl, **c.** NaBr, and **d.** NaI. The blue circles, red squares and green triangles show the results for NaXHe compounds in the $P6_3/mmc$, $P\bar{6}m2$ and $I4_1md$ space groups. The reaction enthalpy, $\Delta H$, is given in eV per formula unit (f.u.).

Our crystal structure searches found three low-energy structures with $P6_3/mmc$, $P\bar{6}m2$, and $I4_1md$ space groups for all four NaXHe compounds (Supplementary Section S1). Below 50 GPa, $I4_1md$ NaClHe is predicted to have the lowest enthalpy, whereas the $P\bar{6}m2$ and $P6_3/mmc$ structures are notably less stable (Figure 1b). In the pressure range that NaClHe is thermodynamically preferred with respect to the salt and noble gas solid, the enthalpies of the three structures are very close, resulting from their structural similarity, as described below. At 300 GPa, the three $\Delta H$ values are -0.22, -0.23 and -0.25 eV per f.u. for the $P\bar{6}m2$, $I4_1md$, and $P6_3/mmc$ structures, respectively. As shown in Figures 1c and 1d, NaBrHe and NaIHe exhibit behavior similar to that of NaClHe, although the corresponding values differ considerably. For NaBrHe, the $I4_1md$ phase is predicted to be preferred below 118 GPa, and the $P6_3/mmc$ above. For NaIHe, the enthalpies of the three candidate structures are virtually indistinguishable below 50 GPa, while above this pressure the enthalpies of the two most stable phases, $P\bar{6}m2$ and $P6_3/mmc$, are significantly lower than that of the $I4_1md$ phase. The dispersion interaction would not have significant on the reaction enthalpy, see Supplementary Section S2. The dynamic stabilities of the NaXHe, X= Cl, Br and I, phases were confirmed by phonon calculations at 50 and 300 GPa (Supplementary Section S3). At these pressures, all three ($P\bar{6}m2$, $P6_3/mmc$ and $I4_1md$) phases were dynamically stable for NaBrHe and NaIHe, while for NaClHe only the phonon band structures of $P\bar{6}m2$ and $P6_3/mmc$ showed no imaginary frequencies. The dynamic stability of the NaFHe phases is discussed in the Supplementary Materials (Supplementary Section S3).

**Crystal structure features of the NaX and NaXHe compounds**

The three NaX phases of concern are the $Fm\bar{3}m$ (B1 phase, rock salt or NaCl structure), $Pm\bar{3}m$ (B2 phase, CsCl structure) and *Cmcm* (B33 phase, CrB structure) (Figure 2a-c). Experiments have shown that both NaF and NaCl undergo a $Fm\bar{3}m \rightarrow Pm\bar{3}m$ phase transition at around 28[36] and 27 GPa[37], respectively. Our first principles calculations (Supplementary Section S4) are in good agreement with experiments and previous theoretical studies,[38,39] predicting these transitions to occur at 30 and 28 GPa, respectively. Unlike NaF and NaCl, NaBr and NaI have both been observed, and calculated, to undergo a transformation from the $Fm\bar{3}m$ to the *Cmcm* structure (parameters given in Supplementary Section S1) between 30 - 40 GPa.[40–42] Our computations predict this transformation to occur at similar pressures for both NaBr and NaI (26 and 22 GPa, respectively). The NaI *Cmcm* structure was experimentally observed to be stable up to 204 GPa.[43] To fully confirm the phases adopted by NaBr and NaI under high pressure, we conducted crystal structure search calculations, and found that NaBr and NaI prefer the *Cmcm* structure up to at least 300 GPa (Supplementary Section S4). Based on these results, the B1 phase was considered as the reference state for all the NaX compounds at 0 GPa, while the B2 and B33 structures were used as references for NaF/NaCl and NaBr/NaI, respectively, above 50 GPa.

All three He-inserted candidate structures ($P6_3/mmc$, $P\bar{6}m2$, $I4_1md$) can be considered as assemblies of Na-$X_6$ trigonal prism building-blocks (Figure 2d-f), where $Na^+$ occupies the center of a trigonal prism formed by 6 $X^-$, paired with a second trigonal prism centered by a He atom. Both the $P\bar{6}m2$ and the $P6_3/mmc$ structures have the $C_3$ axes of the prisms oriented along the *c* direction of the hexagonal unit cells, differing by how these trigonal units are arranged. In $P\bar{6}m2$, $Na^+$ and He occupy the same site along the *c* direction (Figure 2e), while in $P6_3/mmc$, they alternate along the *c* direction (Figure 2d), due to the presence of an additional screw axis. Due to the structural similarity between the two phases, the $c/a$ lattice constant ratio calculated for $P6_3/mmc$ is virtually twice the value calculated for $P\bar{6}m2$. Notably, the alternating ABAB stacking of the NaXHe units in $P6_3/mmc$ phase enhances its stability, especially at higher pressures (Figure 1). This enhanced stability arises from reduced electrostatic repulsion in this structure. Interestingly, the $P6_3/mmc$ structure has some common traits with to the high-pressure phase of sodium Na-hP4,[44] a known high-pressure electride.[45] The Na-X framework coincides with the Na-hP4 network, while the He atoms lie in the same sites occupied by the localized electrons of the electride. This analogy between He-bearing compounds and high-pressure electrides suggests a promising approach for predicting novel systems within these two classes of materials. Finally, the $I4_1md$ phase differs from both the $P6_3/mmc$ and the $P\bar{6}m2$ structures as it possesses two distinct rows of trigonal prisms, with their $C_3$ axis oriented along either $a$ or $b$ unit cell directions (Figure 2f). Similar to the $P\bar{6}m2$ phase, $I4_1md$ has 1-D pores filled with He, along two different directions.

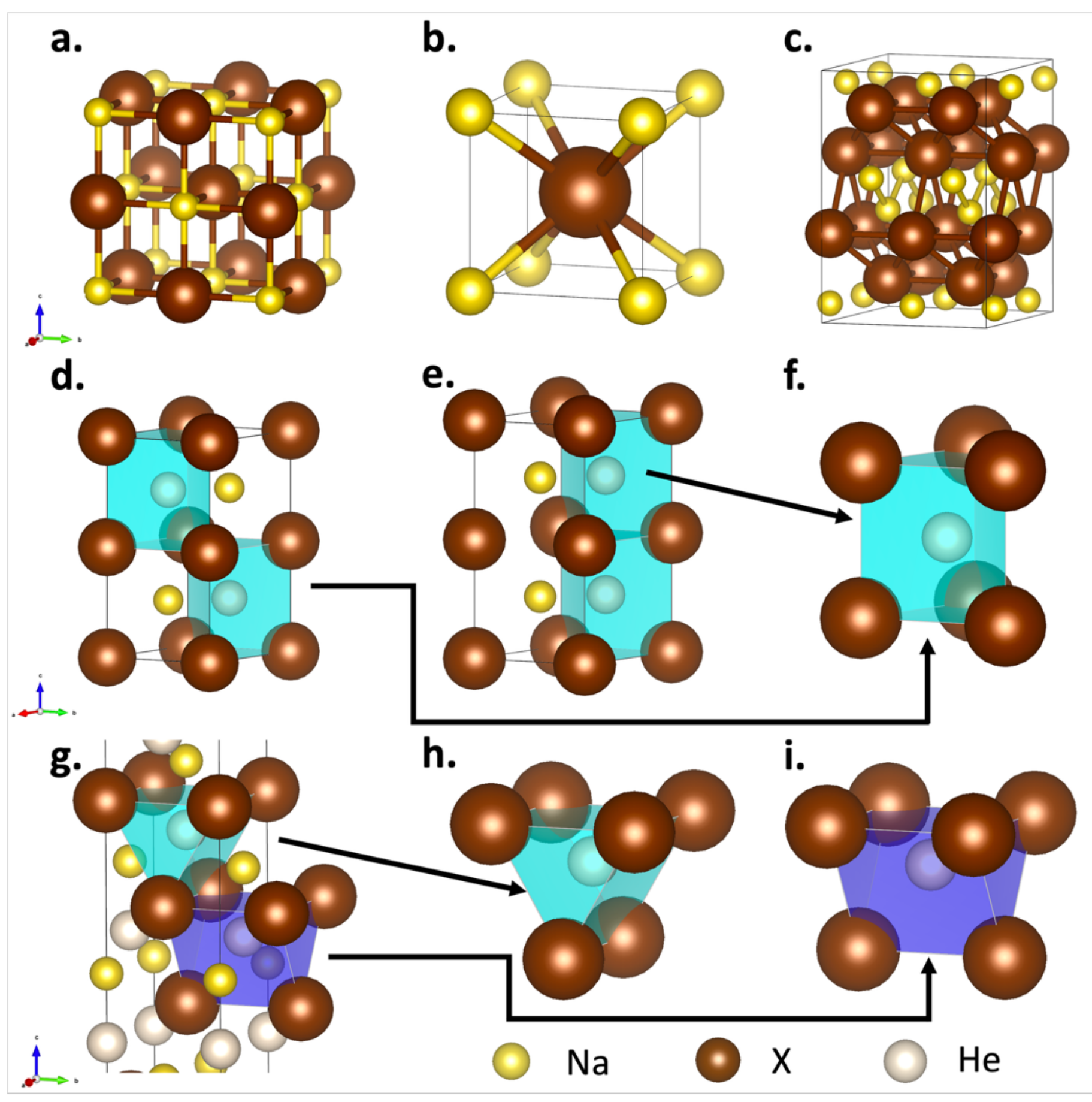

**Figure 2. Selected NaX and NaXHe structures**. **a-c** are the structures of NaX in the $Fm\bar{3}m$, $Pm\bar{3}m$ and $Cmcm$ space group, respectively. **d.** and **e.** are the He inserted NaX structure in the $P6_3/mmc$ **and** $P\bar{6}m2$ space group, respectively. **f.** represents the repeating $HeX_6$ unit and its stacking orientation. **g.** is the NaXHe structure in the $I4_1md$ space group, which has two kinds of repeating $HeX_6$ units that differ in the stacking orientation, highlighted by different colors as shown in **h** and **i.**

It is important to emphasize that, in addition to size, the ratio between the lengths of the two distinct types of edges (base or lateral) comprising the $NaX_6$ trigonal prism is significantly influenced by the size of the halogen atoms (See Supplementary Section S5). The ratio between the lateral to base edges of these triangular prisms varies from 0.75 (F), to 1.01 (Cl), 1.05 (Br), and 1.09 (I) at 300 GPa. Interestingly, both the volume and shape of the prism are important for accommodating He atoms, which can only be inserted into halides whose prisms exhibit a lateral-to-base ratio close to 1. The details can be found in Supplementary Section S5. The corresponding mechanism will be explained and discussed in the next section.

## Discussion

**The change of internal energies and volumes** Under pressure, the enthalpy, which depends on both the pressure-volume ($PV$) term and the internal energy ($E$) term is the key thermodynamic quantity determining stability. We therefore split the reaction enthalpy of He insertion into two terms, $\Delta H = P\Delta V + \Delta E$ (Figure 3). Our calculations show that the changes in both the internal energy and in the volume make significant contributions to the reaction enthalpy. At lower pressures, the insertion of He into most NaX compounds except NaF reduce the volume, resulting in negative $P\Delta V$ terms that stabilize NaXHe. This effect becomes weaker with increasing pressure. In the cass of NaBr and NaI, $P\Delta V$ starts to increase when the pressure goes beyond 100 GPa and becomes positive when the pressure is higher than 300 GPa for NaBr and 150 GPa for NaI. On the other hand, the insertion of He into NaF increases the volume, which is not favored under pressure.

In contrast to volume change, the insertion of He increases the internal energy for all NaX compounds at low pressures. However, for the heavier halides $\Delta E$ decreases steadily with increasing pressure and becomes negative by 242 GPa for NaCl, 73 GPa for NaBr, and 63 GPa for NaI. Thus, both the volume reduction and the internal energy change contribute to the favorable insertion of He into NaX compounds under pressure. Below 250 GPa for NaCl, 100 GPa for NaBr, and 50 GPa for NaI, the volume reduction effects dominate, while at higher pressures, the major driving force of forming He inserted compounds arises in the internal energy change. In what follows, we investigate the origin of the volume and internal energy reduction for He insertions by examining the geometry changes and the electronic structure changes, respectively.

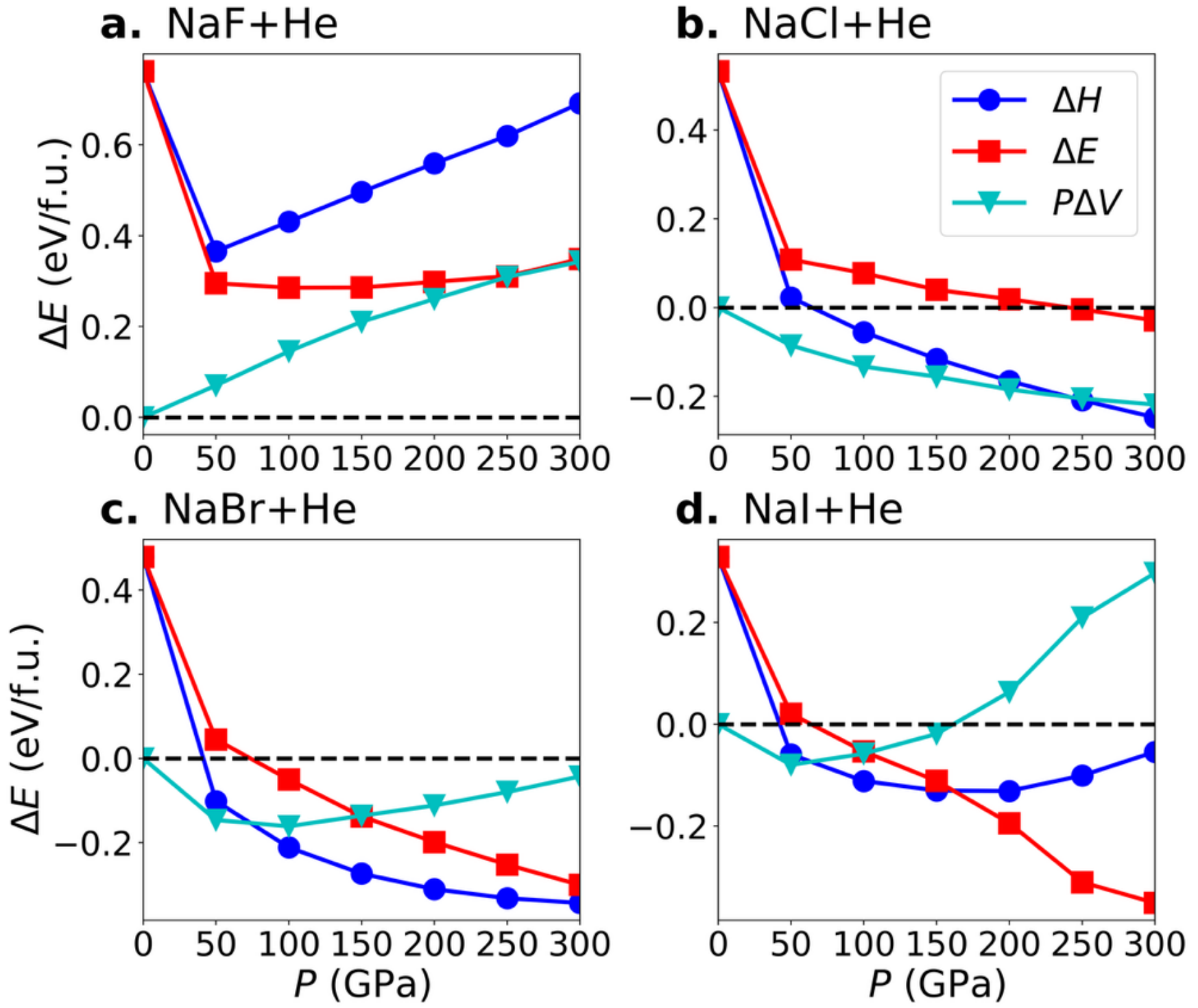

**Figure 3. The reaction enthalpies, change in internal energies, and change in the *PV* term in He insertion reactions. a.** with NaF, **b.** with NaCl, **c.** with NaBr and **d.** with NaI. The blue circles, the red squares, and the green triangles show results of $\Delta H$, $\Delta E$, and $P\Delta V$.

### The Absence of Halogen-Helium bonds

First, we examine whether the volume and internal energy changes in NaX+He reactions are caused by chemical interactions of the host lattice with the inserted He atoms. To investigate this several electronic structure analysis tools were employed, including calculations of the Electron Localization Function (ELF), the Bader charges based upon the Quantum Theory of Atoms in Molecules (QTAIM), the Projected Density of States (PDOS), and the Integrated Crystal Orbital Hamiltonian Population (ICOHP). None of these methods indicate significant chemical interactions between He and the neighboring atoms that could explain the exceptional stability of the NaXHe compounds under pressure. ELF is a real space analysis method used to identify covalent chemical bonds, lone pairs, or atomic shells, which are typically characterized by high ELF values (>0.5).[46,47] The ELF cross sections of $P6_3/mmc$ NaBrHe and NaIHe at 200 GPa along the (110) plane (Figure 4a, b) are characteristic of the closed atomic shells of $Na^+$, $X^-$ and He, and do not show any signatures of covalent bonding. Similar conclusions can be reached for NaFHe and NaClHe (Supplementary Section S6). The computed Bader charges[48] on the He atom are found to be -0.02, -0.05, -0.07 and -0.10, for NaFHe, NaClHe, NaBrHe and NaIHe, respectively, at 200 GPa, which are too small to suggest any significant charge transfer between He and the neighboring atoms, which could result in ionic interactions (Supplementary Section S5). The PDOS results at 200 GPa for all He-inserted compounds (Figure 4c, d) indicate that the He 1s states lie approximately 15 eV below the Fermi level, confirming that they do not participate significantly in direct chemical interactions. Finally, the calculated He–X ICOHP values are too small to be regarded as covalent bonds within the limits of the accuracy of the method. In fact, these He–X ICOHP values are very similar to those computed for Na–X interactions, which are predominantly ionic (Table S7 in Supplementary Section S6). Overall, there is no evidence of direct ionic or covalent bonding between He atoms and the neighboring $Na^+$ and $X^-$ ions that could account for the stability observed upon the insertion of He into NaX compounds. Despite the absence of chemical bond formation, the He insertion indeed affects the electronic structure. The corresponding analyses can be found in Supplementary Section S7.

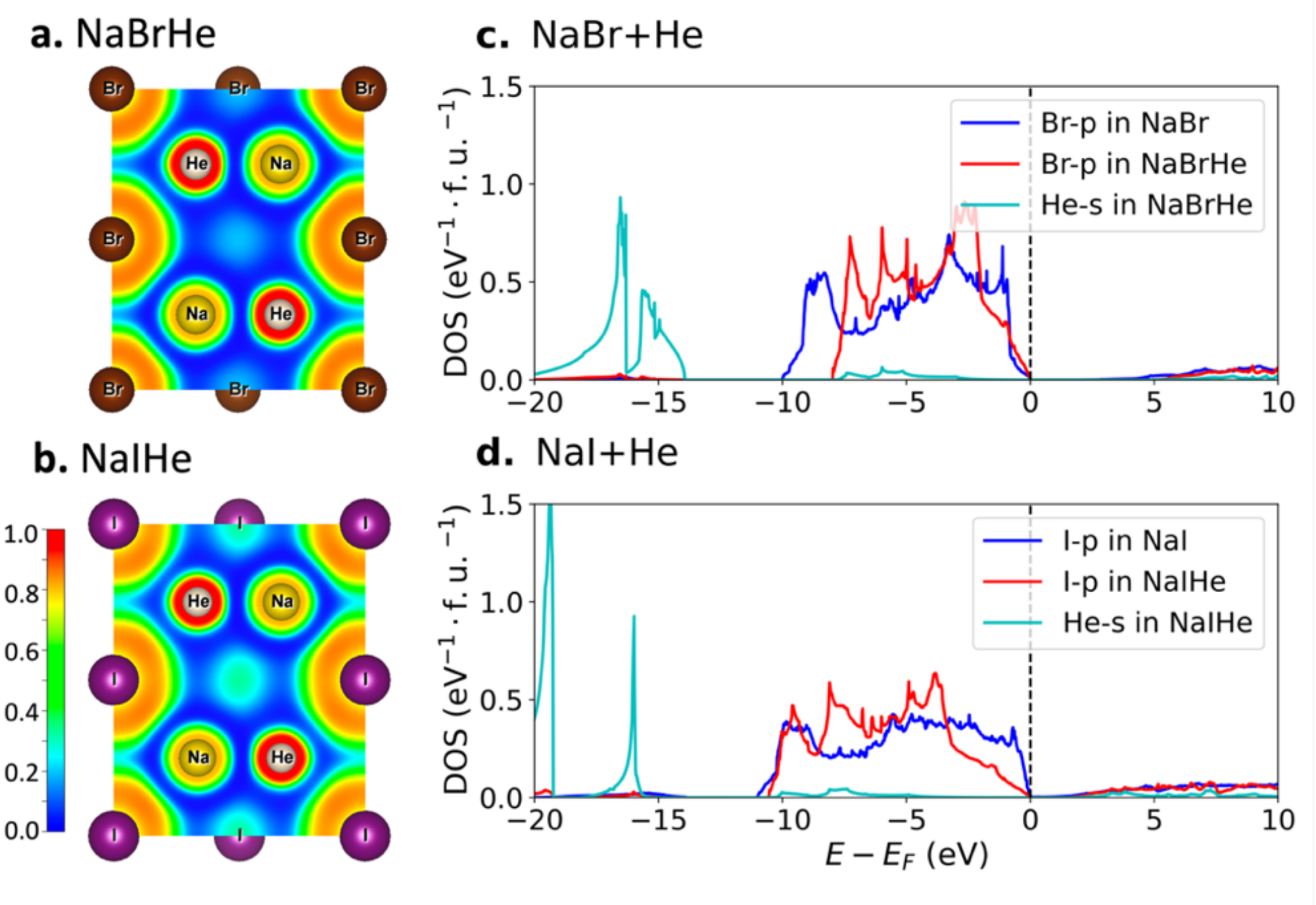

**Figure 4. Electronic structure, including ELF and PDOS, for selected compounds.** The electron localization function (ELF) of **a.** NaBrHe and **b.** NaIHe in the $P6_3/mmc$ structure. The atomic projected densities of states (PDOS) at 200 GPa of: **c.** Br-p states in NaBr in the ***Cmcm*** structure, and Br-p and He-s states of NaBrHe in the $P6_3/mmc$ structure, and **d.** I-p states of NaI in the *Cmcm* structure, and I-p and He-s states of NaIHe in the $P6_3/mmc$ structure. The zero of energy is the top of the conduction band.

**Volume disproportionation**

The difference in the volume between the reactants and products in a solid state reaction depends on many factors, including the rearrangement of the atoms, the redistribution of the charge density and the compressibility of the various atom types. For reactions involving ionic compounds, the charge redistribution might change the size of the ions, causing the volume change to be very hard to predict. However, for the He insertion into NaX compounds, the charge redistribution and the corresponding radii changes are insignificant, and the volume change is mainly caused by the insertion of He atoms and rearrangements of $Na^+$ and $X^-$ ions. In the following analysis, we describe the mechanism underlying the reduction in the volume change and explain why a higher ratio of ionic radii $r_{X^-}/r_{Na^+}$ leads to a more pronounced total volume reduction upon He insertion.

The formation of an NaXHe compound can be viewed as the insertion of He into a peculiar NaX

lattice. At ambient pressure, all sodium halides adopt the B1 ($Fm\bar{3}m$) structure, where $Na^+$ cations and $X^-$ anions each form an FCC lattice, occupying the octahedral interstitial sites of one another. The tetrahedral interstitials (Figure 5a) of the two FCC lattices are identical in the NaCl structure. Each interstitial is surrounded by 4 $Na^+$ and 4 $X^-$ ions with equal distance to its center. If NaClHe is formed by inserting He into NaCl in the B1 structure, only half of the interstitials will be filled with He. This structure is not optimal for He insertion because the empty interstitials are the same size as the filled ones. High-pressure NaX structures, such as those with $Pm\bar{3}m$ and $Cmcm$ space group (B2 and B33 phases), do not fit He insertions for similar reasons. In the $Pm\bar{3}m$ structure, each NaX unit hosts 3 octahedral and 12 tetrahedral interstitials.

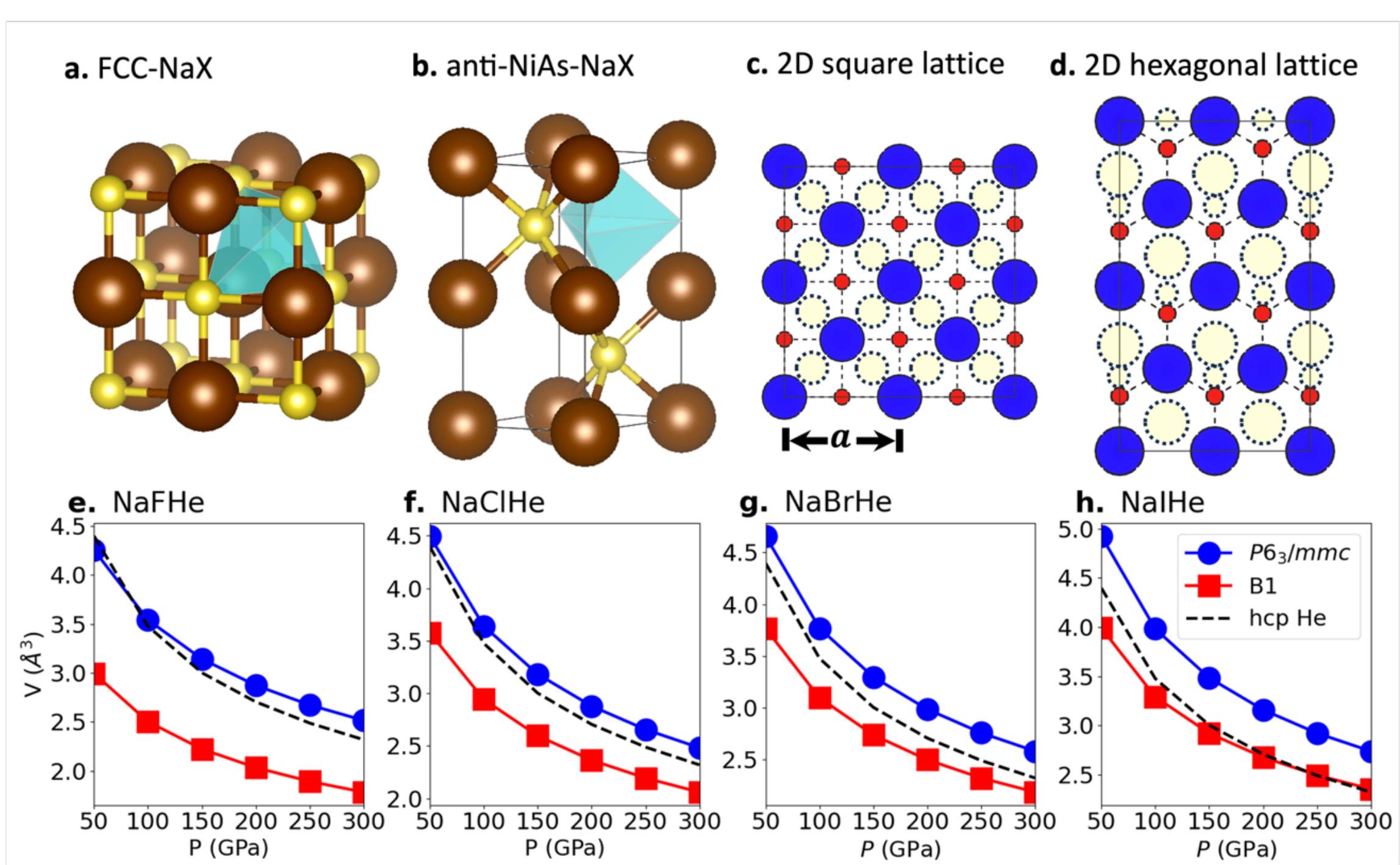

**Figure 5. Volume disproportionation in He insertion reactions with NaX compounds. a.** and **b.** represents the interstitial sites in FCC-NaX and anti-NiAs-NaX. **c.** and **d.** 2D schematic diagrams for FCC with the equal interstitial sites and NiAs structure with distinctly unequal size interstitial sites. **e-h** are the comparison of sizes between He atom and interstitial sites in FCC (B1) and anti-NiAs ($P6_3/mmc$) structures for NaFHe, NaClHe, NaBrHe and NaIHe, respectively.

In contrast to NaX structures, the $Na^+$ and $X^-$ ions in stable NaXHe structures such as $P6_3/mmc$, form an anti-NiAs structure in which $X^-$ anions form a simple hexagonal lattice, and $Na^+$ cations fill half of its octahedral interstitials, see Figure 5b. This structure possesses two types of interstitials, including unoccupied octahedral sites in the $X^-$ lattice and the tetrahedral sites formed by three $X^-$ and one $Na^+$ ions. The volume of the He-occupied octahedral sites is 4 times the volume of the unoccupied tetrahedral sites (Supplementary Section S8). This distinct difference of the interstitial volumes provides sufficient space for the He insertion, while maintaining a low total volume.

Furthermore, the transformation of the NaX lattice during the He insertion also cause an interatomic distance optimization. Going beyond a simple view of volume change, the He insertion is actually determined by the He-Na and He-X interatomic distances. Using FCC structure as an example, the two distances are equal since He atoms occupy the "cubic" interstitials that are the tetrahedral interstitials of both $Na^+$ and $X^-$ FCC lattices. In contrast, when He atoms occupy the prism-center sites in the anti-NiAs NaX, their distances to Na and X atoms are different. For example, in the stable NaClHe compound, the nearest neighbor He-Cl and He-Na distances are 2.543 Å and 1.883 Å. Although He seems to occupy the interstitial sites surrounded by $X^-$ ions, their distances to neighboring $Na^+$ ions are remarkably shorter. Thus, the insertion of He and the formation of NaXHe compounds are more favored for halides with larger $X^-$ anions and larger $r_{X^-}/r_{Na^+}$ ratios.

**The relief of the electrostatic energy** We now investigate why the $\Delta E$ associated with He insertion into NaX (X=Cl, Br, I) decreases under increasing pressure and becomes negative at sufficiently high pressures. The local chemical bonds between He and neighboring atoms are weak and cannot be the cause of the internal energy change. As discussed in our previous study,[32] the internal energy should go up due to He insertion into $A^+B^-$ type ionic compounds, because this insertion would push the $A^+$ and $B^-$ ions with opposite charges and attractive interactions away from each other, thereby increase the electrostatic energy (Rows 1 and 2 in Figure 6a). As we demonstrate in Figure 6a and the following Madelung energy calculations, the size difference between the $A^+$ and $B^-$ ions can explain the driving force for the He insertion. If the anions are much larger than the cations in $A^+B^-$ compounds, the high symmetry structures are not favored under high pressure because they contain large areas of voids (Row 3 in Figure 6a). The reduction of the volume will increase the electrostatic energy since the ions with the same charges are forced to stay closer compared with the high symmetry structures (Row 4 in Figure 6a). Remarkably, the insertion of the He atoms can greatly relieve the so-developed electrostatic stresses resulting in the adoption of low-volume and high-symmetry structures without increasing the electrostatic energies (Row 5 in Figure 6a).

This electrostatic stress relief mechanism is distinctly demonstrated in He insertion in NaX (X=Cl, Br, I) compounds. Increasing pressure stresses these compounds to reduce their volume by transforming them into structures with lower symmetry. NaCl (see Supplementary Section S3), NaBr, and NaI transform into the *Cmcm* structure at pressures of 318, 26, and 22 GPa. In this structure, the halogens form a compact lattice with open channels that accommodate $Na^+$ cations, which remain in close proximity to one another. For example, at 100 GPa, the Na-Na distance in NaI is 2.29 Å in the *Cmcm* structure, which is much smaller than 3.46 Å or 3.07 Å in $Pm\bar{3}m$ and in $Fm\bar{3}m$ structures, respectively. In contrast, the I-I distance is 3.11 Å in the *Cmcm* structure, which is slightly smaller than the 3.46 Å measured in the $Fm\bar{3}m$ structure or slightly larger than the 3.07 Å found in the $Pm\bar{3}m$ structure. Compared with $Fm\bar{3}m$ and $Pm\bar{3}m$ structures, the Madelung energies of NaBr and NaI in the *Cmcm* structure are significantly higher, and the differences become steadily larger under increasing pressures (Figures 6b and 6c). For example, the Madelung energies of *Cmcm* -NaBr/NaI relative to their $Fm\bar{3}m$ structure are -0.08/-0.04 eV per f.u. at ambient conditions and 1.01/1.12 eV per f.u. at 300 GPa.

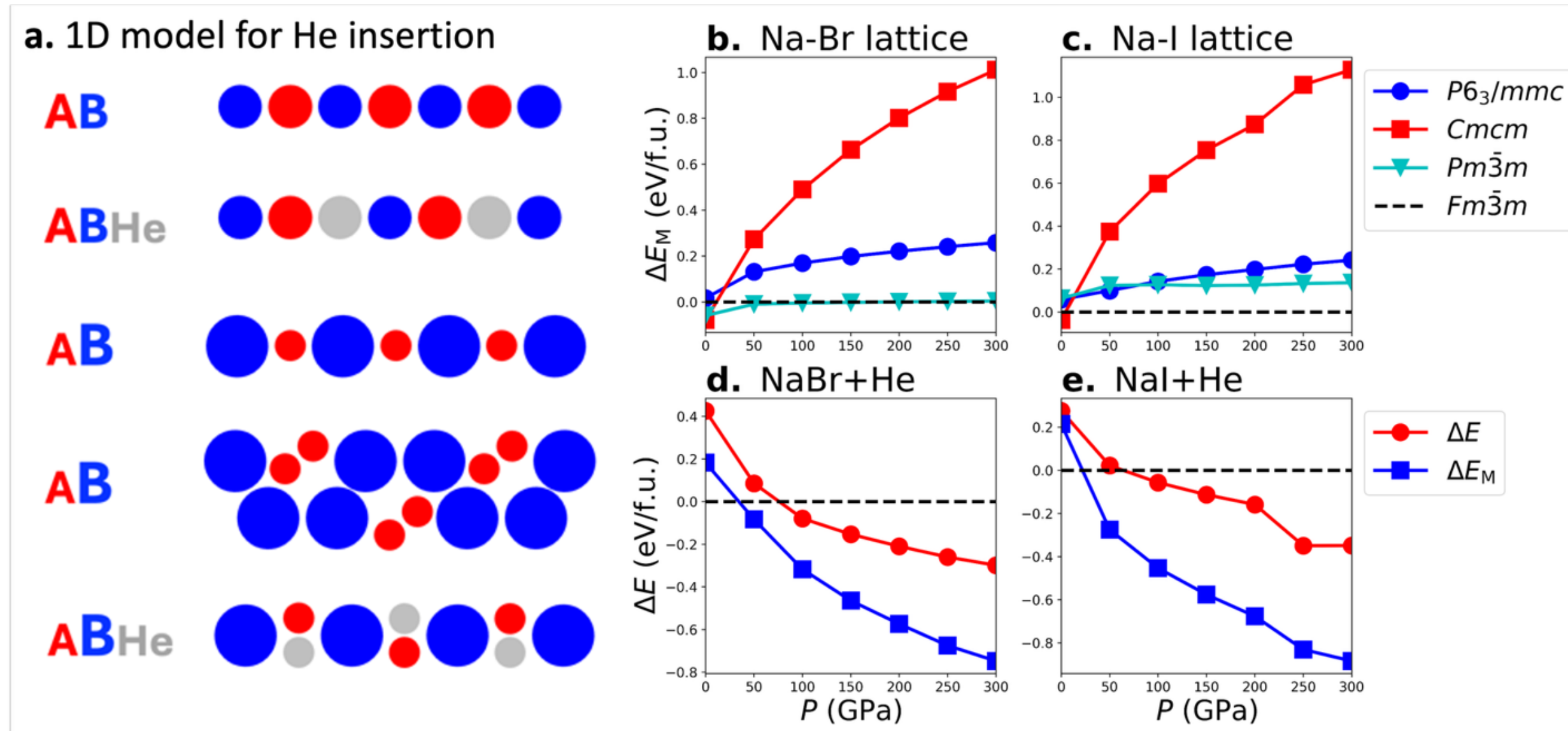

**Figure 6. Electrostatic stress relief mechanism in He insertion reactions with AB-type of ionic compounds. a.** Schematics of He insertion in AB-type ionic compounds. The red, blue and grey circles represent the A+ and B- ions and the He atoms. **b.** and **c.** are the Madelung energy comparison for different Na-Br and Na-I structures. The values are relative to that of the FCC structure. **d**. and **e**. are the difference of Madelung energy and internal energy of He insertion into NaBr and NaI at high pressure.

The strong electrostatic stress in NaX under pressure due to the large size difference between $X^-$ and $Na^+$ ions can be greatly relieved by inserting He atoms. As shown in Figure 6, the Madelung energies within the NaXHe compound in the $P6_3/mmc$ structure are lower than NaX in the *Cmcm* structure. For example, the Madelung energies at 300 GPa are -6.78 and -5.86 eV per f.u. for NaBrHe and NaIHe in the $P6_3/mmc$ structure, much lower than the -6.03 and -4.98 eV for the NaBr and NaI in the *Cmcm* structure. For both compounds, the change in the Madelung energy for the insertion reaction decrease constantly with increasing pressures. It corresponds to the same pressure trend as the internal energy change (Figure 3) but with a larger magnitude. The reduction of the Madelung energy after He insertion can be well explained by the fact that the Na-Na distances become much larger in the inserted structure. For example, the shortest Na-Na distance in NaIHe at 100 GPa is 3.26 Å, much larger than the 2.29 Å in NaI in the *Cmcm* structure. In contrast, the I-I distance, which is also 3.26 Å, is only slightly larger than the I-I distance of 3.11 Å in *Cmcm* NaI. It is worth noticing that $\Delta E_M$ is calculated in our work by treating $Na^+$ and $X^-$ as point charges, which is the reason that their values are larger than $\Delta E$. Other reasons that may cause the difference between $\Delta E_M$ and $\Delta E$ include the bonding changes between atoms, especially the X-X bond change in He insertion reactions. This point will be thoroughly discussed in the next part.

**Bonding Analyses**

Above, we have provided models to help provide intuitive pictures as to why the insertion of He into NaCl, NaBr and NaI is favorable, but into NaCl it is not. While our analysis explicitly compared the pressure-volume terms in the products and reactants, we did not explicitly evaluate the terms involved in the electronic energy contribution to the enthalpy. To better understand the bonding and mechanism of stabilization of NaXHe we performed calculations using atom-centered basis sets on the geometries optimized with planewave calculations. This made it possible to perform an energy decomposition analysis (EDA) - the Ziegler-Rauk variant [49,50]– which decomposes the total interaction energy ($\Delta E_{\mathrm{Int}}$) between two prepared fragments into chemically meaningful terms (see Methods section). The term "prepared" means that the atoms in the moieties that are compared possess the same local coordinates as in the final product.

In EDA, the total interaction energy, $\Delta E_{\text{Int}}$, is decomposed into a Pauli repulsion energy, $\Delta E_{\text{Pauli}}$, which is usually destabilizing; a dispersion energy term, $\Delta E_{\text{Disp}}$, which accounts for van der Waals-like interactions; an electrostatic term, $\Delta E_{\text{Elec}}$, which can quantify the ionic character of an interaction, and finally an orbital term, $\Delta E_{\text{Orb}}$, which accounts for the covalent character of the total interaction. Often, the Pauli repulsion energy and the electrostatic energy are coupled together in a term named the steric interactions, $\Delta E_{\text{Steric}}$, which combines all the charge-charge components in a chemically intuitive term.

The total interaction energy, $\Delta E_{\text{Int}}$, calculated by EDA should not be confused with the $\Delta E$ term in the calculation of the enthalpy of formation, $\Delta H$. In our case, we will apply this methodology to investigate once again the reaction NaXHe – (NaX[He] + [NaX]He) at 200 GPa, and the computed terms are reported in Table 1. Therefore, the analysis will calculate the interaction energy upon insertion of He in the NaX structure, using the geometry of the optimized NaXHe structure (but with one of the sublattices removed). Notice that all the interaction energies, $\Delta E_{\text{Int}}$, are positive, since they account only for the electronic energy associated with the reaction (they do not consider the $PV$ contribution; See Methods Section for further details).

The magnitudes of the terms contributing to the total interaction energy are similar for NaFHe, NaClHe and NaBrHe, despite the fact that NaFHe is, as expected (Figure 1), the most unstable system. The various energy terms calculated from the energy decomposition are almost identical for NaClHe and NaBrHe, apart for $\Delta E_{\text{Orb}}$, which is more negative for NaBrHe. Nonetheless, for these three systems, $\Delta E_{\text{Elec}}$ remains the main stabilizing term, meaning that the insertion of [NaX]He in NaX[He] lowers the electrostatic energy. In Supplementary Section S6, we showed that no appreciable charge transfer occurs during the He intercalation process. Therefore, the large electrostatic stabilization must result from a different mechanism. If we further decompose the steric interaction, defined in the EDA as the sum of $\Delta E_{\text{Pauli}}$ + $\Delta E_{\text{Elec}}$ (see Methods Section), into its fundamental energy components (kinetic, Coulomb and exchange-correlation, See Table S8), it can be seen that the Coulomb energy, accounting for the repulsions/attractions between charges of the same/opposite type, is negative and the main stabilizing component, partially counterbalancing the increase of the kinetic energy, caused by stuffing He in the NaX[He] cavity and increasing the Pauli repulsion. The large, negative, Coulomb energy in the steric interaction mirrors the mechanism proposed by Liu *et al*. for the stabilization upon He insertion into $AB_2$- or $A_2B$-type ionic compounds, where He screens the electrostatic repulsion between ions of the same type, in our case $Na^+$.

NaIHe experiences a similar, but not identical, stabilization process. In this case the main stabilizing contribution is of a different nature. The $\Delta E_{\text{Pauli}}$ is 1.5-2 eV/f.u. larger than for any other species at the same pressure, explained by the larger size of the iodide anion. However, the total $\Delta E_{\text{Int}}$ is also the lowest for any of the species comprising the NaXHe series, and the most stabilizing term turns out to be the orbital relaxation ($\Delta E_{\text{Orb}}$), not the electrostatic interactions, making NaIHe a rather covalent compound. Therefore, for NaIHe we propose that the He not only acts as a buffer between the halides and the three sodium ions lying on the same plane, reducing the electron-electron repulsion, and therefore, making the Coulomb energy more negative (Table S8), but the electrons of the iodide atoms are also able to repolarize and relax, reducing the global kinetic energy (See Table S8) and lowering the orbital energy, and so, the total interaction energy (Table 1).

**Table 1.** Energy decomposition analysis, calculated for the reaction NaXHe – (NaX[He] + [NaX]He) at 200 GPa, in eV/f.u., at the BP86-D4/TZP level of theory.

| **BP86-D4/TZP** | **NaFHe** *Pnma* | **NaClHe** $P6_3/mmc$ | **NaBrHe** $P6_3/mmc$ | **NaIHe** $P6_3/mmc$ |
|---|---|---|---|---|
| $\Delta E_{\text{Pauli}}$ | 5.63 | 5.99 | 6.00 | 7.56 |
| $\Delta E_{\text{Disp}}$ | -0.11 | -0.15 | -0.14 | -0.16 |
| $\Delta E_{\text{Elec}}$ | -1.70 | -1.86 | -1.89 | -2.11 |
| $\Delta E_{Steric}$ | 3.93 | 4.13 | 4.11 | 5.45 |
| $\Delta E_{\text{Orb}}$ | -0.65 | -0.93 | -1.05 | -2.63 |
| $\Delta E_{\text{Int}}$ | 3.17 | 3.06 | 2.92 | 2.67 |

Why is He needed for this repolarization? To answer to this question let's have a look at the -ICOHPs calculated with plane-waves (Table 2) calculated for the stable NaXHe and NaX[He] (X = Cl, Br, and I).

The three halides behave similarly upon the insertion of He: the He-X covalent energy of interaction is almost zero, and similar to those calculated in other systems at similar pressures;[33] while Na-X interactions are slightly strengthened.

On the other hand, the X-X interaction along the *a* and *b* crystallographic directions (Supplementary Section S5), seems to be reinforced upon intake of He in the bromide and chloride system, but has almost no effect on the iodide.

However, the X-X' interaction, which lies along the *c* direction, is notably reinforced in NaIHe, but less in NaBrHe and almost unaffected in NaClHe. Surprisingly, the longer the halide-halide interaction (relative to X-X and X-X') the larger is the covalent stabilization. It was shown by Blokker *et al.*[51] that, analyzing carbon-halogen bonds, C-I is weaker that C-F due to an high Pauli repulsion, generated by the large size of iodine. Therefore, increasing the distance between two big atoms forced to be too close to each other, might actually be beneficial for strengthening the chemical bond.

Interestingly, He seems to increase the covalent character of NaIHe by supporting the I-I' bonds. The strengthening of these interactions upon intake of the noble gas is likely another side effect of the large increment of Pauli repulsions (Table 1). The presence of He in cavity forces the electrons of iodine to repolarize towards the only available free space, *i.e.*, between the iodides I-I', concentrating the charge and strengthening the covalent bonds. This effect is particularly evident in NaIHe compared to the other cases due to the highly polarizability of the iodide ions. Therefore, He is also able to increase the strength of covalent bonds by forcing a repolarization of the charge density. We can see this polarization effect in the three systems by calculating the charge density difference (CDD) isosurfaces. In Figure 7, it is shown how the charge density reorganizes upon intake of He, highlighting a substantial depletion in the cavity and simultaneously concentrating at both the He position and along the X-X' interaction. Moreover, the higher polarizability of iodine here is evident in the shifting of the charge concentration loci towards the center of the I-I' interaction.

**Table 2.** The crystal orbital Hamilton population integrated to the Fermi level (-ICOHP), in [eV/bond], of the interactions involving halogen atoms in $P6_3/mmc$ NaXHe and NaX[He] (X= Cl, Br and I).

| -ICOHP | NaClHe | NaCl[He] | NaBrHe | NaBr[He] | NaIHe | NaI[He] |
|---|---|---|---|---|---|---|

| [eV\bond] | | | | | | |
|---|---|---|---|---|---|---|
| Na-X | 0.91 | 0.84 | 1.05 | 0.93 | 1.12 | 1.04 |
| X-X | 0.10 | 0.07 | 0.09 | 0.01 | 0.34 | 0.35 |
| X-X' | 0.05 | 0.02 | 0.21 | 0.13 | 0.44 | 0.34 |
| X-He | -0.01 | \ | 0.00 | \ | 0.01 | \ |

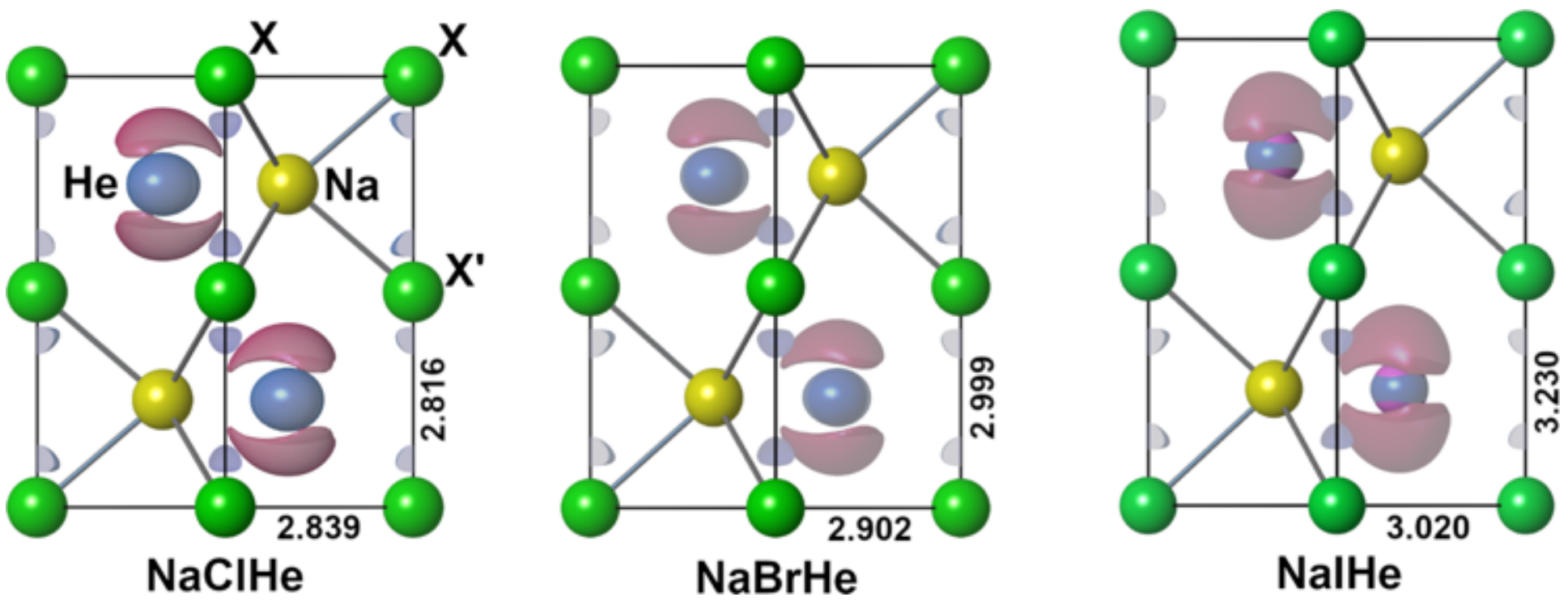

**Figure 7**. Charge density (ρ) difference isosurface (isovalues = 0.0035), calculated as ρ(NaXHe) – [ρ(NaX[He]) + ρ([NaX]He)] in the *P6₃/mmc* NaXHe systems (X= Cl, Br and I) at 200 GPa. Color-code: blue = increase, red = depletion.

**Conclusions**

In summary, our large-scale crystal structure searches based on density functional total energy and geometry relaxations discovered that He atoms can be inserted into NaX (X=Cl, Br, I) crystals and form stable NaXHe compounds under high pressure. At first sight, this result opposes the previous proposal that He atoms cannot be inserted into AB-type ionic compounds because this process will increase the electrostatic energy. In-depth geometry and electronic structure analyses reveals that this unusual He insertion reaction is driven by a key factor missing in the previous theory, namely the large $r_{X^-}/r_{Na^+}$ ratios, which can lead to the reduction of both volume and internal energy. The reductions in volumes can be realized by a volume disproportionation mechanism, through which NaX compounds accommodate He atoms in structures containing interstitial sites with very different volumes. Especially, the He insertion sites have very different interatomic distances toward the neighboring Na+ and X-, which is advantageous while $r_{X^-}/r_{Na^+}$ ratio is large. Furthermore, the large $r_{X^-}/r_{Na^+}$ ratios force the NaX to adopt structures with short Na-Na distances and high electrostatic energies under high pressure. The insertion of He atoms can greatly relieve such electrostatic stress in NaX and lower the total internal energy. Our work reveals that He, being one of the most stable elements, may have very rich chemistry under high pressure that is governed not by the local chemical bonds but by long-range electrostatic interactions. It also suggests rich chemistry of He in the interior of Earth and giant planets, where minerals consist of ions with different charges and volumes.

## Method

**Structure search** The CALYPSO (Crystal structure AnaLYsis by Particle Swarm Optimization)[52–55] package was used for crystal structure prediction searches. The CALYPSO code was employed to find the most stable structures containing a mixture of He and sodium halides (NaF, NaCl, NaBr and NaI) in a 1:1 ratio at 100 GPa, 200 GPa and 300 GPa. The unit cells were chosen to contain 2 - 4 formula units (f.u.'s). To ensure the PSO structure searches were likely to find the global minimum structure each search procedure was set to create 30 generations of 50 structures, and it was repeated several times. In each generation, 60% of the structures were generated by the PSO algorithm and 40% of the structures were generated randomly.

**Relaxation, enthalpy, and phonon calculations** The stability and electronic structure computations were carried out using density functional theory (DFT) as implemented in the Vienna Ab-initio Simulation Package (VASP).[56] We used the generalized gradient approximation for the exchange-correlation functional within the framework of Perdew-Burke-Ernzerhof.[57] These structures were also examined with the optB88-vdW exchange-correlation functional to account for dispersion interactions.[58–60] The electron-ion interactions were described by the projector augmented wave method within the frozen core approximation.[61,62] The following electronic configurations were treated explicitly: $1s^2$, $2s^22p^63s^1$, $2s^22p^5$, $3s^23p^5$, $4s^24p^5$ and $5s^25p^5$ for He, Na, F, Cl, Br and I, respectively. A 1400 eV plane wave cut-off energy was employed for all calculations, coupled with a Brillouin zone sampling scheme with Monkhorst-Pack k-point meshes and a spacing between points of no more than $2\pi \times 0.05 Å^{-1}$ in reciprocal space. Dynamical stability of the predicted NaXHe phases was determined using VASP, combined with the PHONOPY package[63] in the harmonic approximation~~, on supercells of size 3x3x2 for the $P6_3/mmc$ phases and 3x2x2 for the *Pnma* phase~~.

**Madelung energy** The Madelung energy is calculated through the Ewald sum rule. In this method, the point charge density distribution $\rho_i(\mathrm{r}) = q_i\delta(\mathrm{r})$ on each atomic site is split into two terms by subtracting then adding a Gaussian function, and one can have $\rho_i(\mathrm{r}) = q_i\delta(\mathrm{r}) - q_iG_\sigma(\mathrm{r}) + q_iG_\sigma(\mathrm{r})$. It is convenient to define the density as the addition of the two parts: $\rho_i^S(\mathrm{r}) = q_i\delta(\mathrm{r}) - q_iG_\sigma(\mathrm{r})$ and $\rho_i^L(\mathrm{r}) = q_iG_\sigma(\mathrm{r})$. The potential of one site generated can be written as $\phi_i(\mathrm{r}) = \phi_i^S(\mathrm{r}) + \phi_i^L(\mathrm{r})$, where $\phi_i^S(\mathrm{r}) = \frac{q_i}{4\pi\epsilon_0}\int\frac{\delta(\mathrm{r})-G_\sigma(\mathrm{r})}{|\mathrm{r}-\mathrm{r}'|}d^3r$ and $\phi_i^L(\mathrm{r}) = \frac{q_i}{4\pi\epsilon_0}\int\frac{G_\sigma(\mathrm{r})}{|\mathrm{r}-\mathrm{r}'|}d^3r$ are the short and long range component of the Coulomb potential involved with an individual point charge density distribution $\rho_i(\mathrm{r})$. Based the Ewald sum rule, the Madelung energy of each unit cell can be written as $E_M = \frac{1}{2}\sum_i^j q_i\,\phi_{[i]}^S(\mathrm{r}) + \frac{1}{2}\sum_i^j q_i\,\phi_{[i]}^L(\mathrm{r})$. The sum in second term can be further calculate in Fourier transformation in reciprocal space because of the long-range property. The calculate method in implemented in a homemade python script. Our code agrees with Madelung energy computed by VESTA, see Supplementary Section S11.

**Bader charge** Chemically meaningful information can be obtained from the topological analysis of the electron density via the Quantum Theory of Atoms in Molecules, developed by Richard Bader.[64] In this analysis, the atomic boundaries in molecules or crystals are defined by the surface zero-flux condition of the electron density calculated for each atom in the system.[65] The region of space defined by the zero-flux condition is called an atomic basin, which can be used to integrate the electronic charge associated with one atom.[66] In Bader's analysis, the difference between the nuclear charge Z and the number of electrons integrated in the atomic basin is called a Bader charge. In this work, we used the Bader charge analysis code[48] to analyze the electron transfer between atoms.

**Integrated Crystal Orbital Hamiltonian Population** For analysis of the chemical bonds strength from an orbital point of view, we calculated the Crystal Orbital Hamiltonian Population (COHP),[67,68] as well as the integrated COHP (ICOHP) to the Fermi level, using the LOBSTER package.[69,70] The value of –ICOHP of selected atom pairs can be used to estimate the strength of the covalent interaction between those atoms.[68,71]

**Bonding Analyses** The topological analyses of the electron density via the Quantum Theory of Atoms in Molecules[64] were carried out to obtain charges. We employed the energy decomposition analysis (EDA) using the Ziegler-Rauk formalism[50,72] implemented for periodic structures within the BAND code[73], which used atom-centered basis sets, at the BP86-D4/TZP[74–76] level. The analysis was performed on the structures optimized with VASP at 200 GPa. In EDA, the total interaction energy, $\Delta E_{\text{Int}}$, is decomposed into a Pauli repulsion energy, $\Delta E_{\text{Pauli}}$, which is usually destabilizing; a dispersion energy term, $\Delta E_{\text{Disp}}$, which accounts for van der Waals-like interactions; an electrostatic term, $\Delta E_{\text{Elec}}$, which can quantify the ionic character of an interaction, and finally an orbital term, $\Delta E_{\text{Orb}}$, which accounts for the covalent character of the total interaction. Often, the Pauli repulsion energy and the electrostatic energy are coupled together in a term named the steric interactions, $\Delta E_{\text{Steric}}$, which combines all the charge-charge components in a chemically intuitive term. The total interaction energy, $\Delta E_{\text{Int}}$, calculated by EDA should not be confused with the $\Delta E$ term in the calculation of the enthalpy of formation, $\Delta H$. In the first case, the electronic energy of interaction is evaluated upon the reaction between two moieties possessing the same geometry as in the final product. In the second, $\Delta H$ is evaluated from the isolated reactants at their ground states (*e.g.*, the crystalline hcp He or the rock salt NaCl), which is composed by the thermodynamic terms $\Delta E$ and $P\Delta V$.

**Acknowledgements**

M.M. and A.P. acknowledge the support of the DoD Research and Education Program for Historically Black Colleges and Universities and Minority-Serving Institutions (HBCU/MI) Basic Research Funding under grant No. W911NF2310232., as well as the support of National Science Foundation (NSF) funds DMR 1848141 and OAC 2117956. M.M. also acknowledges the support of ACF PRF 59249-UNI6, the Camille and Henry Dreyfus Foundation, and California State University Research, Scholarship and Creative Activity (RSCA) awards. Z.L. and D. Y. acknowledge the National Natural Science Foundation of China (NSFC) for grants under No. 12004045 and No. 22173013, respectively. We acknowledge the Center for Matter at Atomic Pressures (CMAP), a National Science Foundation (NSF) Physics Frontier Center, under Award No. PHY-2020249 (S.R. and E.Z.), and the U.S. Department of Energy, Office of Science, Fusion Energy Sciences Award no. DE-SC0020340 (E.Z.). K.H. is thankful to the U.S. Department of Energy, National Nuclear Security Administration, through the Capital-DOE Alliance Center under Cooperative Agreement DE-NA0003975 for financial support. Some calculations were also performed at the Center for Computational Research at SUNY Buffalo (https://hdl.handle.net/10477/79221).

**Author contributions**

M.M. and E.Z. designed, coordinated, and guided the research. Z.L. and S.R. contribute equally to this study. M.M. and Z.L. proposed the volume disproportionation and size-disparity driven Madelung relief mechanism. Z.L. conducted the crystal structure searches and the volume and Madelung energy analyses. K.P.H. did the chemical pressure calculations and analyses. S.R. and E.Z. proposed and conducted the bond analyses and energy decomposition analysis. All authors were involved in the data analysis and discussion of the results. M.M., Z.L., S.R., E.Z., A.H., K.P.H., D.Y. wrote and revised the manuscript together.

**Additional information**

Supplementary Information accompanies this paper at https:xxx.xxx.xxx

# Non-local Chemistry Driven by Cation-Anion Size Disparity in Helium Inserted Compounds under High Pressure

*Zhen Liu,[1,2] Stefano Racioppi,[3] Katerina P. Hilleke,[3] Abhiyan Pandit,[4] Shuran Ma,[1] Andreas Hermann,[5] Dadong Yan,[6] Eva Zurek,[3] Mao-sheng Miao[4*]*

1. *School of Physics and Astronomy, Beijing Normal University, Beijing 100875, China*
2. *Key Laboratory of Multiscale Spin Physics (Ministry of Education), Beijing Normal University, Beijing 100875, China*
3. *Department of Chemistry, State University of New York at Buffalo, Buffalo, New York 14260-3000, United States.*
4. *Department of Chemistry and Biochemistry, California State University Northridge, California 91330-8262, USA*
5. *Centre for Science at Extreme Conditions and SUPA, School of Physics and Astronomy, The University of Edinburgh, Edinburgh EH9 3FD, United Kingdom*
6. *School of Physics, Zhejiang University, Hangzhou 310058, China*

**Supplementary Section S1:** The structural parameters of NaXHe and NaX

The parameters of optimized structures of NaXHe (X=F, Cl, Br and I) at 300 GPa are shown in Table S1-S4. In addition, the *Cmcm*- NaBr and NaI structure parameters are shown in Table S5.

**Table S1.** The structural parameters of NaFHe at 300 GPa

| Formula | Space group | Pressure (GPa) | Lattice Parameters (Å, °) | | | Atomic coordinates (Fractional coordinates) | | | |
|---|---|---|---|---|---|---|---|---|---|
| NaFHe | $I4_1md$ | 300 | $a = 2.327$ | $b = 2.327$ | $c = 9.380$ | Na | 0.000 | 0.000 | 0.112 |
| | | | $\alpha = 90$ | $\beta = 90$ | $\gamma = 90$ | F | 0.000 | 0.000 | 0.524 |
| | | | | | | He | 0.000 | 0.000 | -0.045 |
| | $P\bar{6}m2$ | 300 | $a = 2.649$ | $b = 2.649$ | $c = 2.063$ | Na | 0.667 | 0.333 | 0.500 |
| | | | $\alpha = 90$ | $\beta = 90$ | $\gamma = 120$ | F | 0.000 | 0.000 | 0.000 |
| | | | | | | He | 0.333 | 0.667 | 0.500 |
| | $P6_3/mmc$ | 300 | $a = 2.680$ | $b = 2.680$ | $c = 4.017$ | Na | 0.333 | 0.667 | 0.750 |
| | | | $\alpha = 90$ | $\beta = 90$ | $\gamma = 120$ | F | 0.000 | 0.000 | 0.000 |
| | | | | | | He | 0.333 | 0.667 | 0.250 |
| | $Pnma$ | 300 | $a = 4.073$ | $b = 2.653$ | $c = 4.588$ | Na | 0.296 | 0.250 | 0.585 |
| | | | $\alpha = 90$ | $\beta = 90$ | $\gamma = 90$ | F | -0.006 | 0.250 | 0.294 |
| | | | | | | He | 0.711 | 0.250 | 0.587 |

**Table S2**. The structural parameters of NaClHe at 300 GPa.

| Formula | Space group | Pressure (GPa) | Lattice Parameters (Å, °) | | | Atomic coordinates (Fractional coordinates) | | | |
|---|---|---|---|---|---|---|---|---|---|
| NaClHe | $I4_1md$ | 300 | $a = 2.718$ | $b = 2.718$ | $c = 9.359$ | Na | 0.000 | 0.000 | 0.114 |
| | | | $\alpha = 90$ | $\beta = 90$ | $\gamma = 90$ | Cl | 0.000 | 0.000 | 0.530 |
| | | | | | | He | 0.000 | 0.000 | -0.053 |
| | $P\bar{6}m2$ | 300 | $a = 2.699$ | $b = 2.699$ | $c = 2.739$ | Na | 0.667 | 0.333 | 0.500 |
| | | | $\alpha = 90$ | $\beta = 90$ | $\gamma = 120$ | Cl | 0.000 | 0.000 | 0.000 |
| | | | | | | He | 0.333 | 0.667 | 0.500 |
| | $P6_3/mmc$ | 300 | $a = 2.707$ | $b = 2.707$ | $c = 5.446$ | Na | 0.333 | 0.667 | 0.750 |
| | | | $\alpha = 90$ | $\beta = 90$ | $\gamma = 120$ | Cl | 0.000 | 0.000 | 0.000 |
| | | | | | | He | 0.333 | 0.667 | 0.250 |

**Table S3**. The structural parameters of NaBrHe at 300 GPa.

| Formula | Space group | Pressure (GPa) | Lattice Parameters (Å, °) | | | Atomic coordinates (Fractional coordinates) | | | |
|---|---|---|---|---|---|---|---|---|---|
| NaBrHe | $I4_1md$ | 300 | $a = 2.831$ | $b = 2.831$ | $c = 9.591$ | Na | 0.000 | 0.000 | 0.391 |
| | | | $\alpha = 90$ | $\beta = 90$ | $\gamma = 90$ | Br | 0.000 | 0.000 | -0.027 |
| | | | | | | He | 0.000 | 0.000 | 0.558 |
| | $P\bar{6}m2$ | 300 | $a = 2.758$ | $b = 2.758$ | $c = 2.910$ | Na | 0.667 | 0.333 | 0.500 |
| | | | $\alpha = 90$ | $\beta = 90$ | $\gamma = 120$ | Br | 0.000 | 0.000 | 0.000 |
| | | | | | | He | 0.333 | 0.667 | 0.500 |
| | $P6_3/mmc$ | 300 | $a = 2.762$ | $b = 2.762$ | $c = 5.803$ | Na | 0.333 | 0.667 | 0.750 |
| | | | $\alpha = 90$ | $\beta = 90$ | $\gamma = 120$ | Br | 0.000 | 0.000 | 0.000 |
| | | | | | | He | 0.333 | 0.667 | 0.250 |

**Table S4.** The structural parameters of NaIHe at 300 GPa.

| Formula | Space group | Pressure (GPa) | Lattice Parameters (Å, °) | | | Atomic coordinates (Fractional coordinates) | | | |
|---|---|---|---|---|---|---|---|---|---|
| NaIHe | $I4_1md$ | 300 | $a = 2.991$ | $b = 2.991$ | $c = 9.978$ | Na | 0.000 | 0.000 | 0.112 |
| | | | $\alpha = 90$ | $\beta = 90$ | $\gamma = 90$ | I | 0.000 | 0.000 | 0.532 |
| | | | | | | He | 0.000 | 0.000 | -0.053 |
| | $P\bar{6}m2$ | 300 | $a = 2.869$ | $b = 2.869$ | $c = 3.214$ | Na | 0.667 | 0.333 | 0.500 |
| | | | $\alpha = 90$ | $\beta = 90$ | $\gamma = 120$ | I | 0.000 | 0.000 | 0.000 |
| | | | | | | He | 0.333 | 0.667 | 0.500 |
| | $P6_3/mmc$ | 300 | $a = 2.868$ | $b = 2.868$ | $c = 6.252$ | Na | 0.333 | 0.667 | 0.750 |
| | | | $\alpha = 90$ | $\beta = 90$ | $\gamma = 120$ | I | 0.000 | 0.000 | 0.000 |
| | | | | | | He | 0.333 | 0.667 | 0.250 |

**Table S5.** The structural parameters of the B33 phase of NaBr and NaI at 300 GPa

| Formula | Space group | Pressure (GPa) | Lattice Parameters (Å, °) | | | Atomic coordinates (Fractional coordinates) | | | |
|---|---|---|---|---|---|---|---|---|---|
| NaBr | $Cmcm$ | 300 | $a = 2.752$ | $b = 7.976$ | $c = 3.076$ | Na | 0.000 | -0.070 | 0.250 |
| | | | $\alpha = 90$ | $\beta = 90$ | $\gamma = 90$ | Br | 0.000 | 0.654 | 0.250 |
| NaI | $Cmcm$ | 300 | $a = 3.220$ | $b = 8.330$ | $c = 2.950$ | Na | 0.000 | 0.066 | 0.250 |
| | | | $\alpha = 90$ | $\beta = 90$ | $\gamma = 90$ | I | 0.000 | 0.348 | 0.250 |

**Supplementary Section S2**: The formation enthalpy including van der Waals interactions

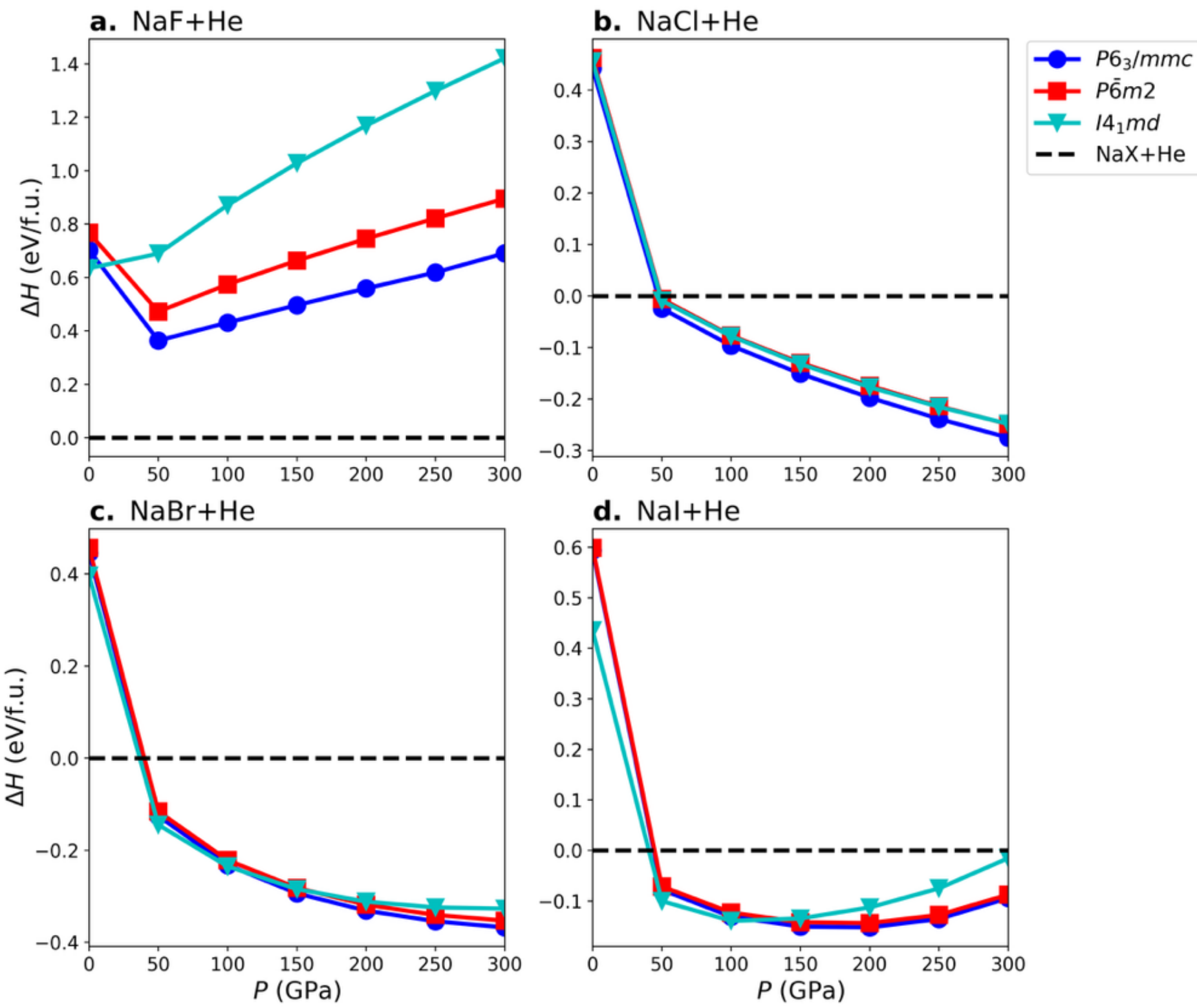

**Figure S1.** The formation enthalpy of **a**. NaF+He, **b**. NaCl+He, **c**. NaBr+He and **d**. NaI+He between 0 and 300 GPa by taking consideration of van der Waals interaction. These structures were reoptimized with the optB88-vdW exchange-correlation functional to account for dispersion interactions.[1–3]

**Supplementary Section S3:** The phonon dispersion relations of NaXHe at high pressure

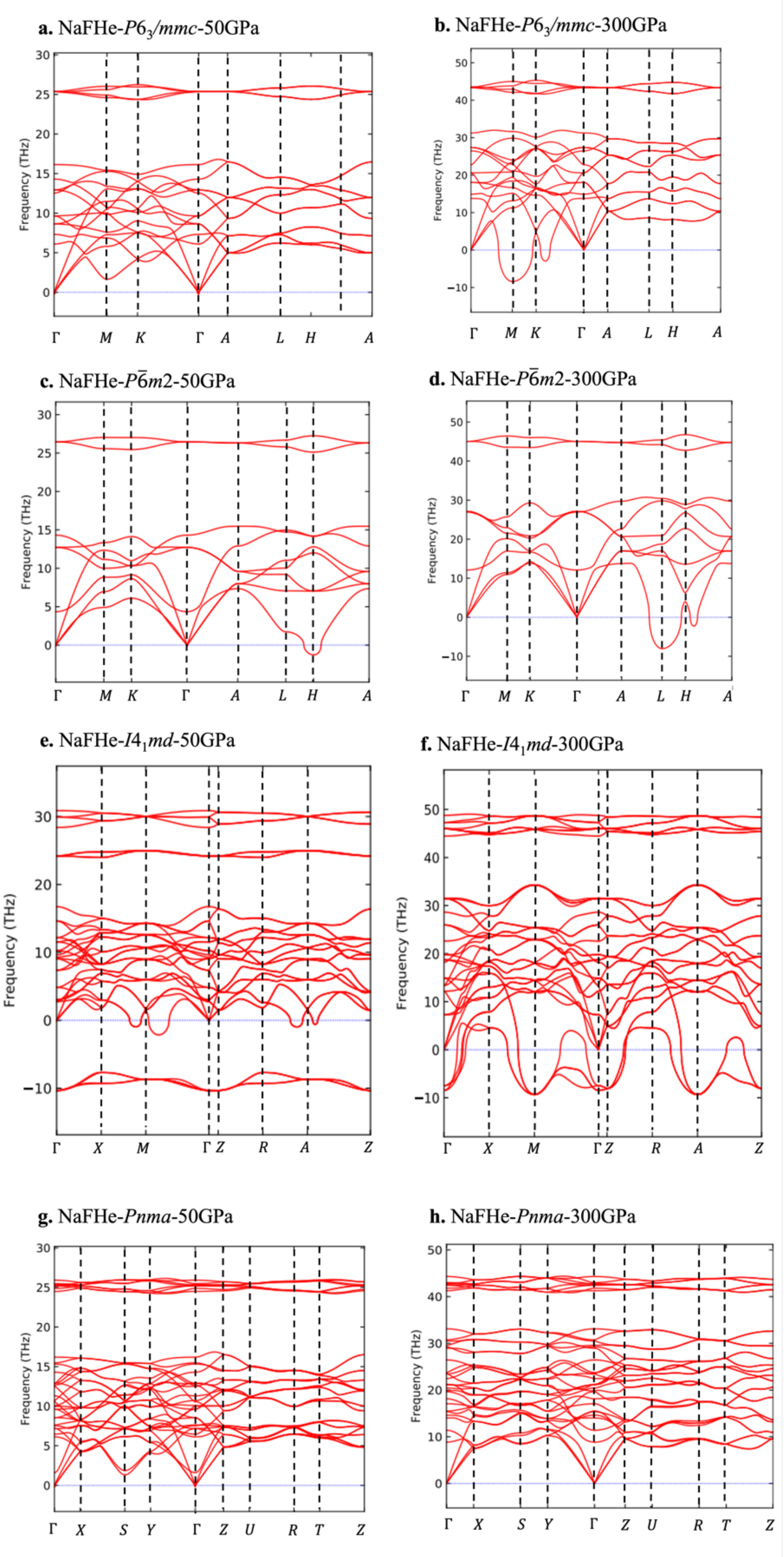

**Figure S2.** The phonon band structure of NaFHe at 50 GPa and 300 GPa

At 300 GPa, NaFHe $P6_3/mmc$ possessed imaginary modes along the M-direction (Figure S2b). Therefore, we have doubled the unit cell of the dynamically unstable structure along the M direction and reoptimized the resulting phase without symmetry constraints, after applying small random atomic displacements. This operation revealed the new *Pnma*-NaFHe phase, see Figure S3. This new structure is not thermodynamically stable, while its phonon dispersion relation shows no imaginary frequency, suggesting that this structure is located at a local minimum on the potential energy surface.

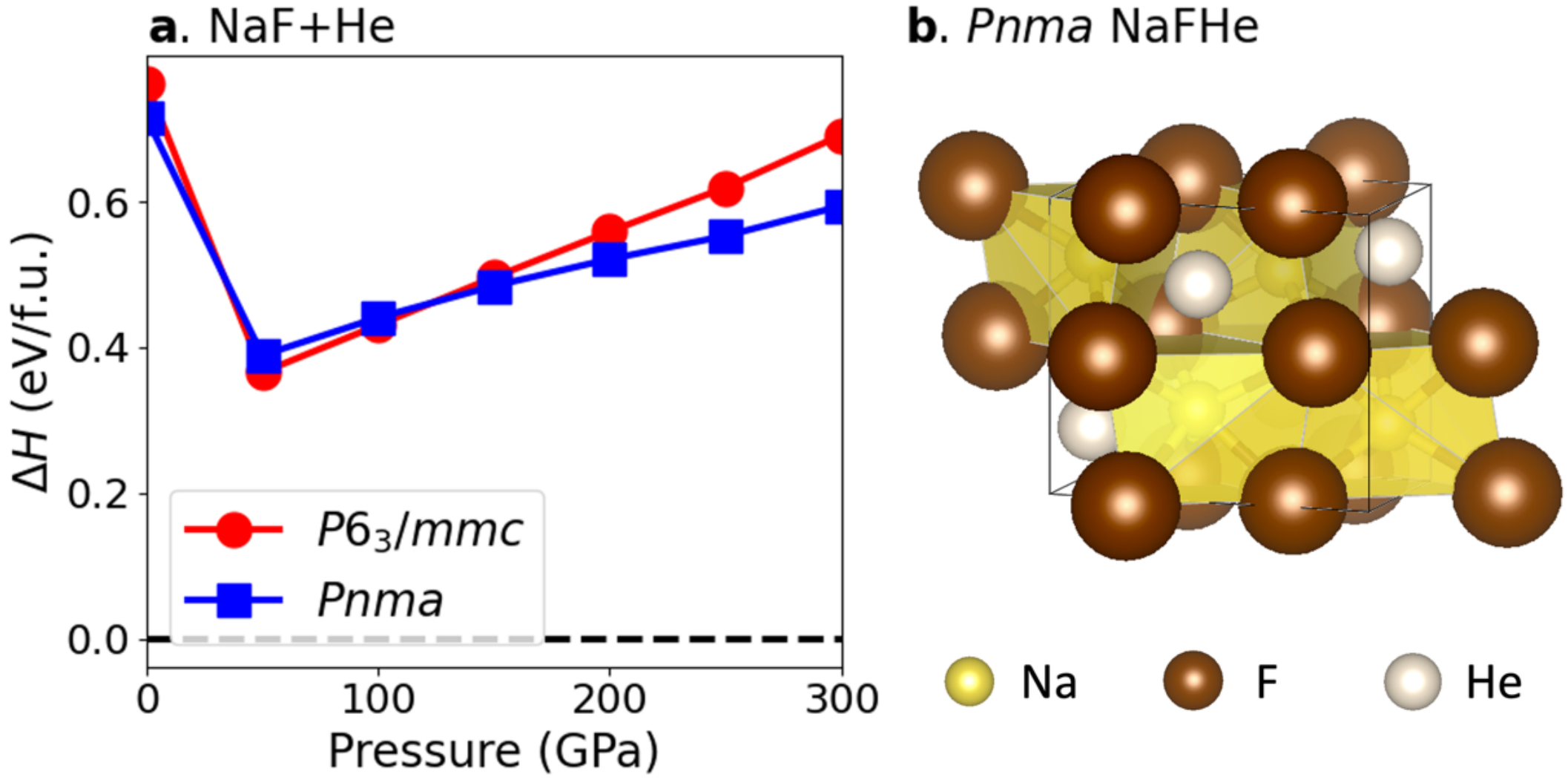

**Figure S3.** The thermodynamic stability and structure of *Pnma*-NaXHe. **a**. the enthalpies of *Pnma*- and $P6_3/mmc$- NaFHe referring to the separate phase of reactant NaF and He. **b**. the structure of *Pnma*-NaFHe

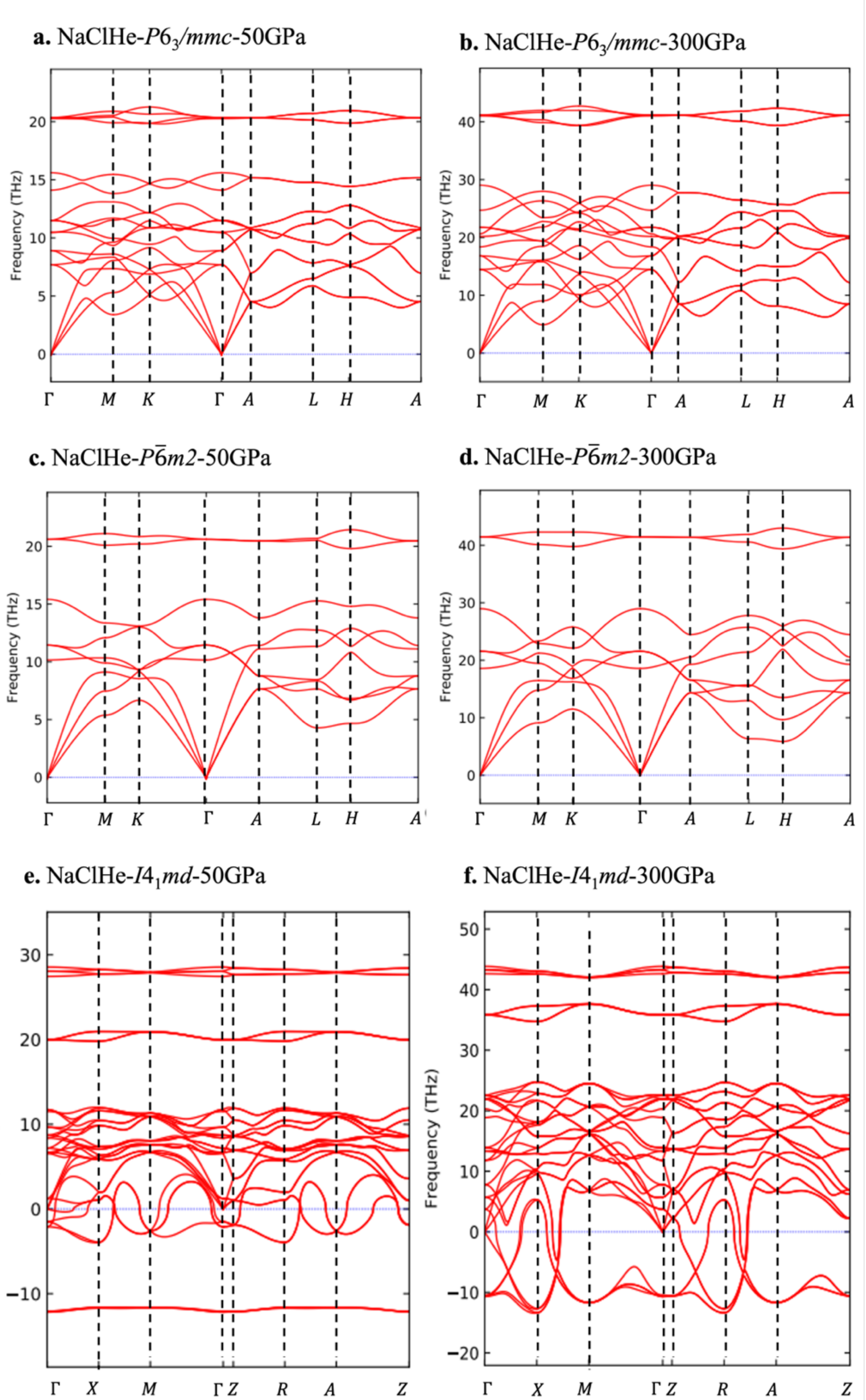

**Figure S4.** The phonon band structure of NaClHe at 50 GPa and 300 GPa

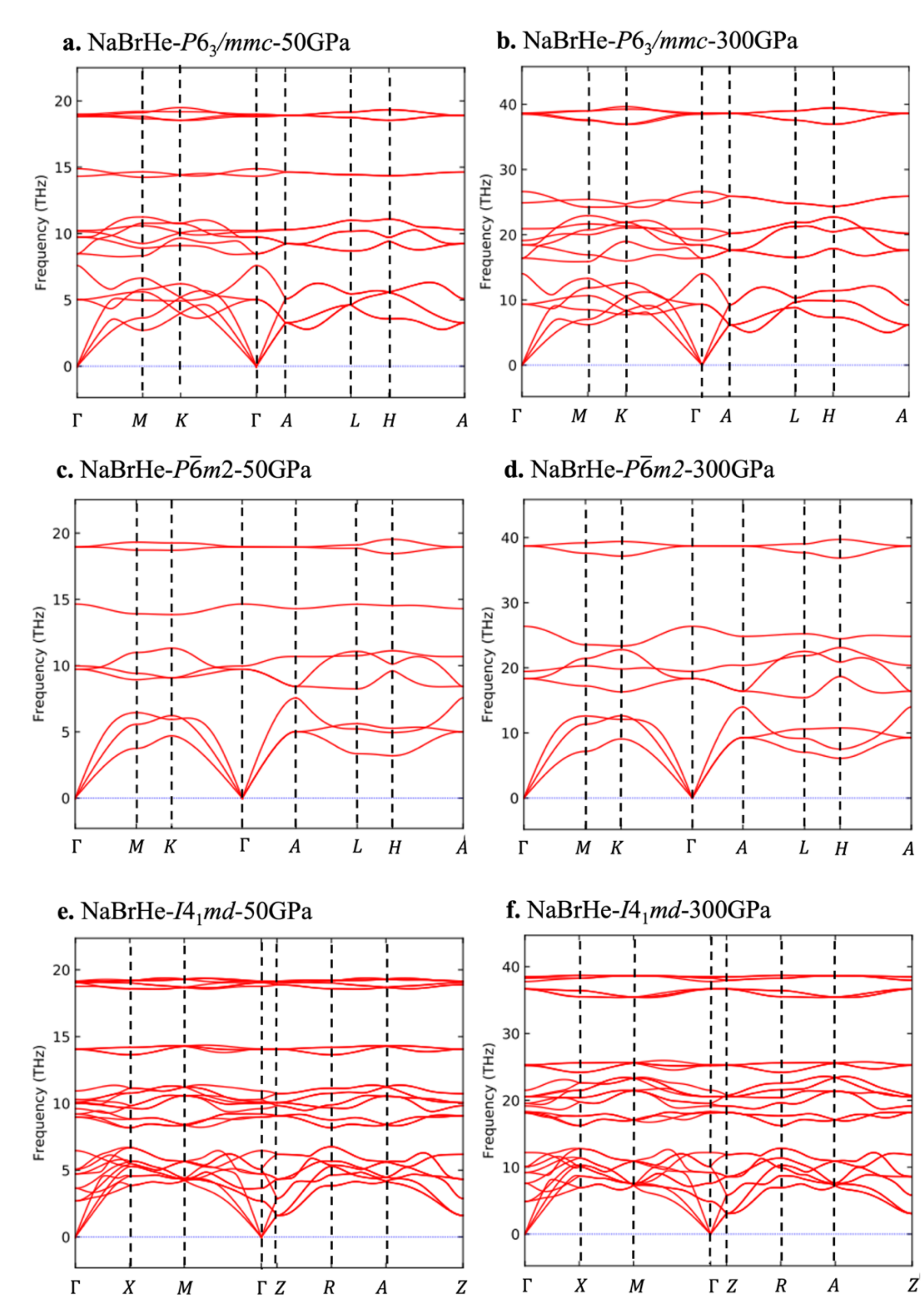

**Figure S5.** The phonon band structure of NaBrHe at 50 GPa and 300 GPa

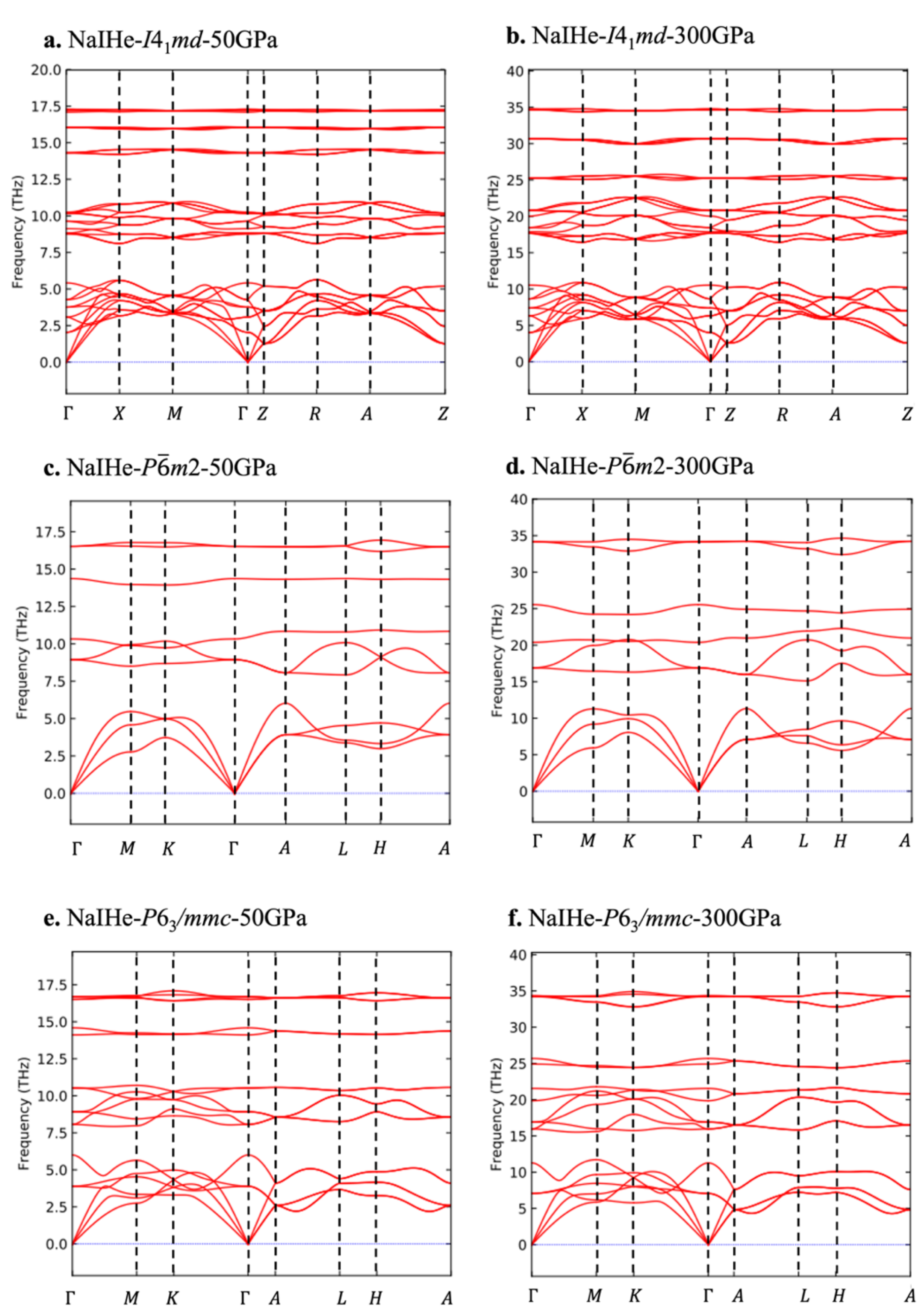

**Figure S6.** The phonon band structure of NaIHe at 50 GPa and 300 GPa

**Supplementary Section S4:** The structural transition of NaX at high pressure

NaF and NaCl both undergo a structural phase transition in which the space group change from $Fm\bar{3}m$ into $Pm\bar{3}m$ (B1-B2 phase transition). To further analyze the structural phases of NaBr and NaI upto 300 GPa, we performed crystal structure search and found a new candidate structure for NaBr and NaI in the *Pmcn* spacegroup, see Figure S7e, which is only ~0.1-0.2 eV per formula unit higher in energy than *Cmcm*-NaBr (NaI). In this structure, each Na ion is coordinated with seven Br/I ions. This Na-$Br_7$ and Na-$I_7$ unit are also observed in *Cmcm*- NaBr and NaI. The difference between *Pmcn*-NaBr (NaI) and *Cmcm*-NaBr (NaI) is the way of stacking these Na-$Br_7$ and Na-$I_7$ unit.

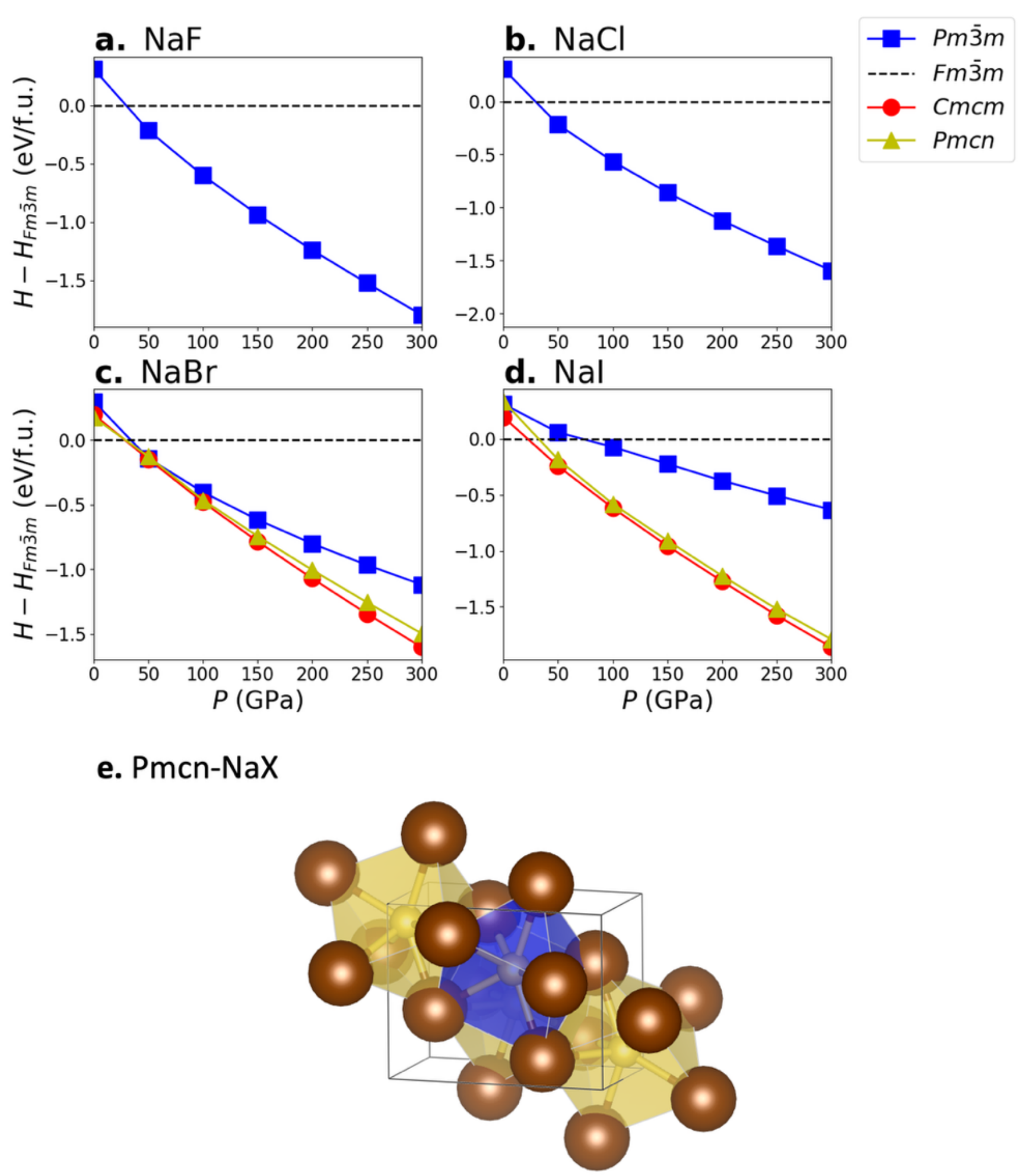

**Figure S7.** The enthalpy comparison of the **a**. NaF, **b**. NaCl, **c**. NaBr and **d**. NaI. **e**. represents the structure of *Pmcn*-NaX.

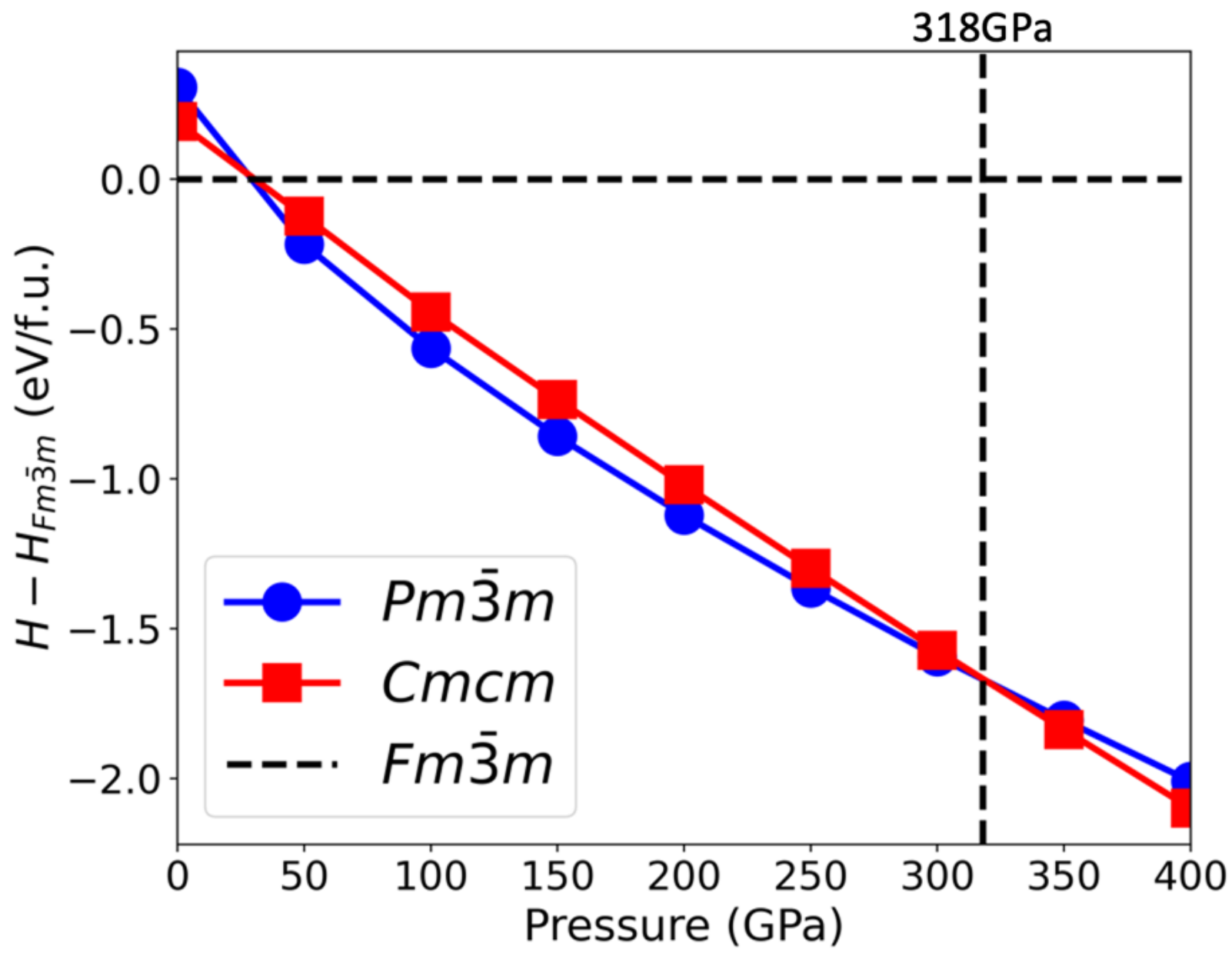

**Figure S8.** The enthalpy comparison of the *Pm*$\bar{3}$*m*-, *Cmcm*-, and *Fm*$\bar{3}$*m*- NaCl from 0 GPa to 400 GPa. The structural transition from *Pm*$\bar{3}$*m*- to *Cmcm*-NaCl happens at 318 GPa.

**Supplementary Section S5:** The deformation of the periodic unit

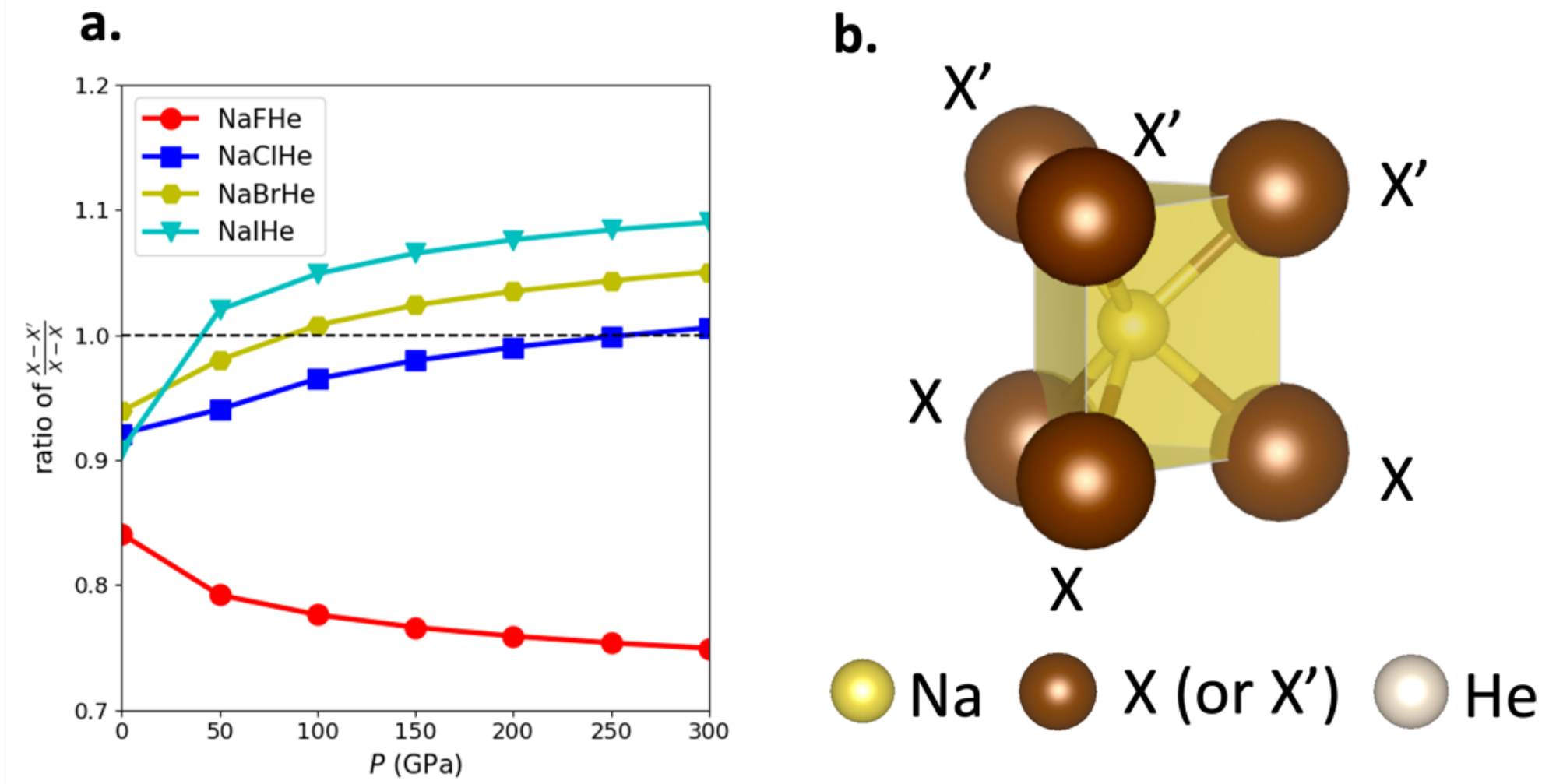

**Figure S9. a.** The (X-X')/(X-X) ratio of the $NaX_6$ block for NaFHe, NaClHe, NaBrHe and NaIHe. **b**. the triangular prism structure of the Na-$X_6$ block.

**Table S6** The X-X' and X-X distance and (X-X')/(X-X) ratio

| 50 GPa | Na(He)-X (Å) | X-X (Å) | X-X' (Å) | (X-X')/(X-X) |
|---|---|---|---|---|
| NaFHe | 2.165 | 3.093 | 2.500 | 0.79 |
| NaClHe | 2.453 | 3.294 | 3.099 | 0.94 |
| NaBrHe | 2.554 | 3.373 | 3.306 | 0.98 |
| NaIHe | 2.707 | 3.513 | 3.584 | 1.02 |

| 100 GPa | Na(He)-X (Å) | X-X (Å) | X-X' (Å) | (X-X')/(X-X) |
|---|---|---|---|---|
| NaFHe | 2.046 | 2.941 | 2.283 | 0.78 |
| NaClHe | 2.309 | 3.070 | 2.962 | 0.96 |
| NaBrHe | 2.403 | 3.136 | 3.161 | 1.01 |
| NaIHe | 2.543 | 3.260 | 3.420 | 1.05 |

| 200 GPa | Na(He)-X (Å) | X-X (Å) | X-X' (Å) | (X-X')/(X-X) |
|---|---|---|---|---|
| NaFHe | 1.920 | 2.779 | 2.110 | 0.76 |
| NaClHe | 2.161 | 2.841 | 2.813 | 0.99 |
| NaBrHe | 2.248 | 2.899 | 3.000 | 1.03 |
| NaIHe | 2.375 | 3.010 | 3.239 | 1.08 |

| 300 GPa | Na(He)-X (Å) | X-X (Å) | X-X' (Å) | (X-X')/(X-X) |
|---|---|---|---|---|
| NaFHe | 1.845 | 2.680 | 2.009 | 0.75 |
| NaClHe | 2.073 | 2.707 | 2.723 | 1.01 |
| NaBrHe | 2.156 | 2.762 | 2.902 | 1.05 |
| NaIHe | 2.277 | 2.868 | 3.126 | 1.09 |

**Supplementary Section S6:** The electronic structure analysis for NaXHe

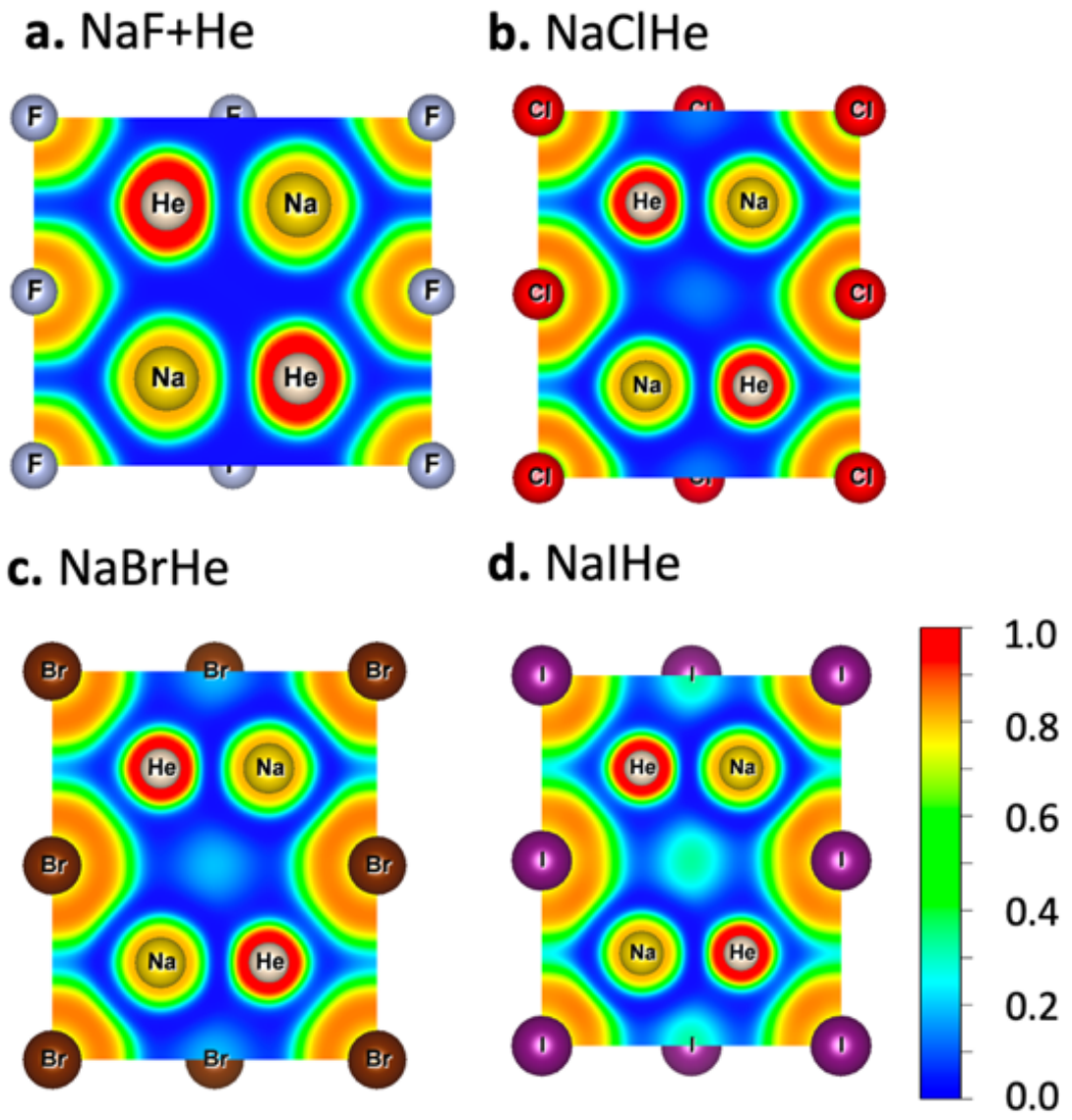

**Figure S10.** The ELF of the He-inserted compounds NaXHe in $P6_3/mmc$ space group cleaved along the (110) face at 200 GPa. **a.** NaFHe **b.** NaClHe **c.** NaBrHe **d.** NaIHe.

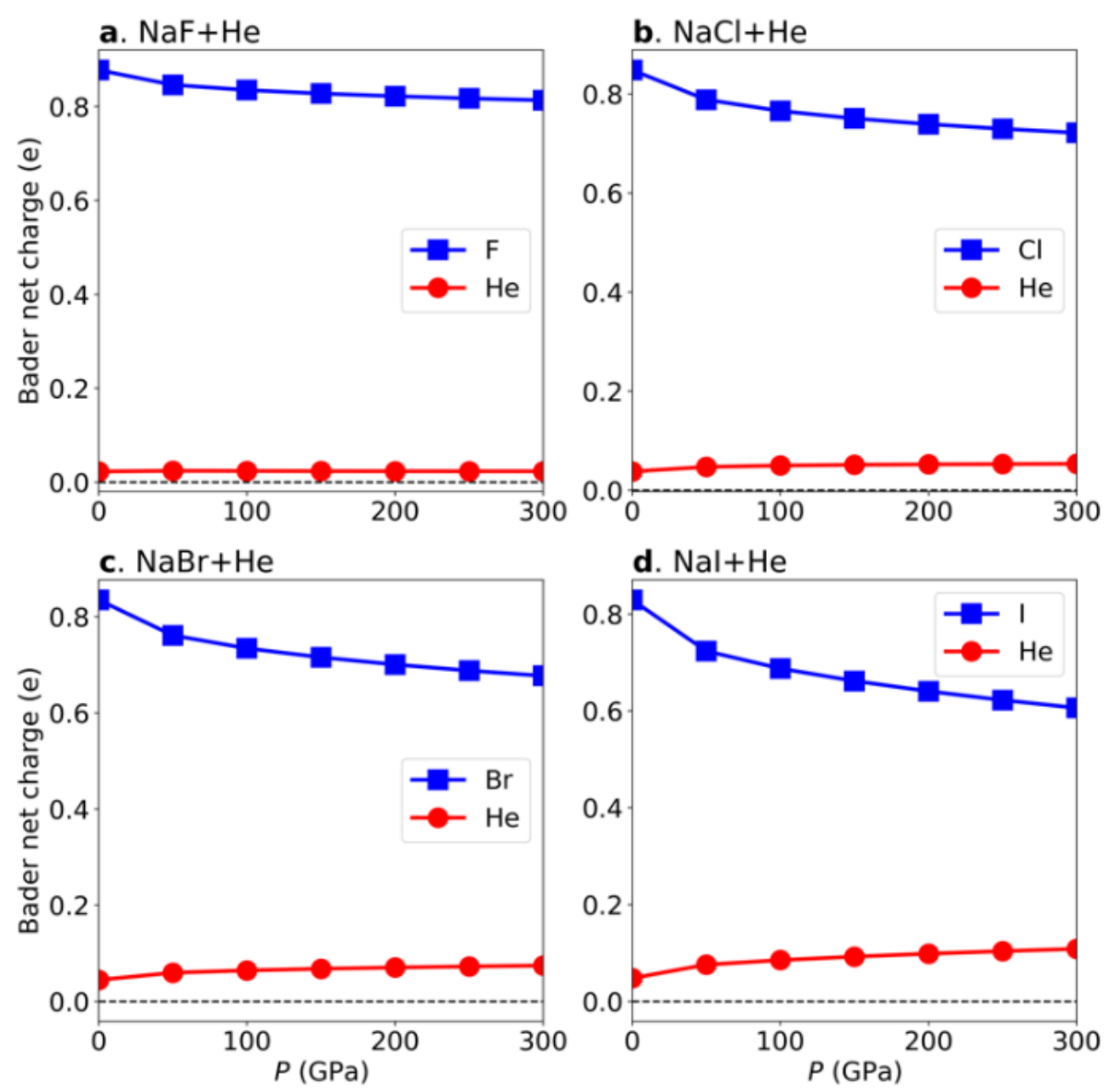

**Figure S11.** The net Bader charge of the halogen anions and He atoms in **a.** NaFHe. **b**. NaClHe, **c**. NaBrHe and **d**. NaIHe.

In Figure S11, the Bader charge of the halogen anions and He atoms at various pressures are displayed. The Bader charges of $F^-$, $Cl^-$, $Br^-$, $I^-$ are close to -1, which refers to the electron transferring from Na to the corresponding halogen atom. The He atom, on the other hand, remains nearly neutral, resulting from its high ionization potential and low electron affinity.

**Table S7.** The ICOHP of Na-X and He-X pairs at 100 GPa, 200 GPa and 300 GPa

| Pressure (GPa) | ICOHP (eV/pair) | NaF | NaFHe | NaCl | NaClHe | NaBr | NaBrHe | NaI | NaIHe |
|---|---|---|---|---|---|---|---|---|---|
| 100 | Na-X | -0.116 | -0.076 | -0.167 | -0.159 | -0.151 | -0.190 | -0.117 | -0.212 |
| | He-X | | 0.005 | | 0.047 | | 0.046 | | 0.055 |
| 200 | Na-X | -0.105 | -0.068 | -0.143 | -0.158 | -0.106 | -0.168 | -0.078 | -0.170 |
| | He-X | | 0.014 | | 0.095 | | 0.093 | | 0.079 |
| 300 | Na-X | -0.090 | -0.059 | -0.119 | -0.106 | -0.071 | -0.124 | -0.072 | -0.129 |
| | He-X | | 0.015 | | 0.113 | | 0.110 | | 0.095 |

**Supplementary Section S7**: The He insertion effect on the electronic structure

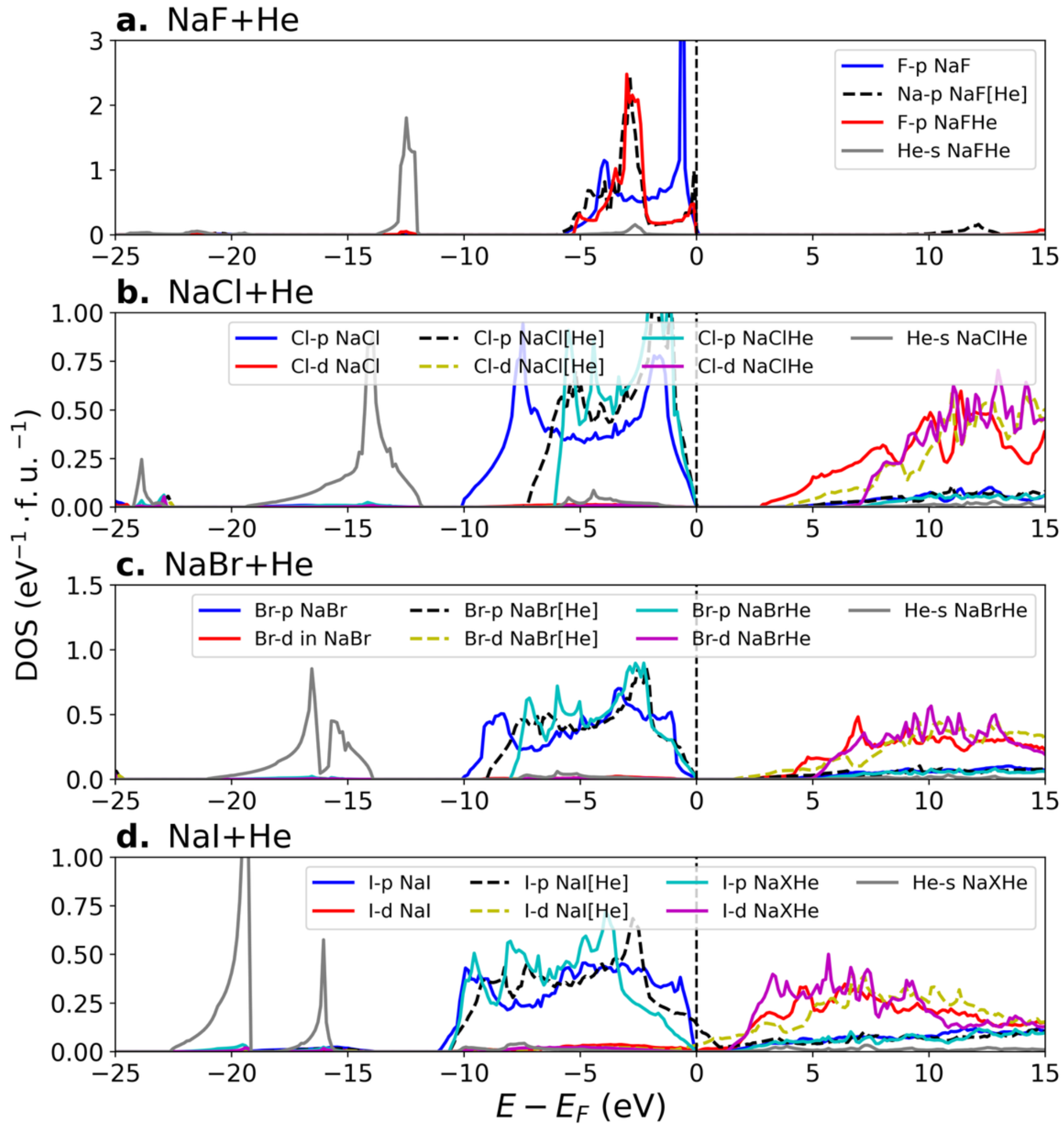

**Figure S12.** The projected density of states for **a**. NaF+He, **b**. NaCl+He, **c**. NaBr+He and **d**. NaI+He in the NaX, NaX[He] and NaXHe structures.

To analyze how He insertion affects the projected density of states (PDOS), an imaginary intermediate Na-X crystal structure, denoted as NaX[He], was constructed by removing the He atom from $P6_3/mmc$ NaXHe but leaving the unit cell and other atomic positions unchanged. The transition from NaX to NaX[He] refers to the Na-X lattice change, while the transition from NaX[He] to NaXHe refers to He insertion. The He $s$ and halogen $p$ PDOS per formula unit of the NaX, NaX[He] and NaXHe systems at 200 GPa are shown in Figure S12. In NaXHe, He $s$-states are fully occupied and located far below the $X^-$ $p$-states. The band widths of the 2$p$-states of F in NaF, NaF[He] and NaFHe are 5.8, 6.0, and 5.3 eV, respectively, at 200

GPa. The change of the Na-F lattice vector and He insertion into NaF[He] to form $P6_3/mmc$ NaFHe does not noticeably perturb the F-2$p$ band width, which is in-line with the chemical hardness and small compressibility of the F- ions. The band widths of the Cl-3$p$ states in $Pm\overline{3}m$- NaCl and NaCl[He] are 10.6 and 7.4 eV (Figure S12b), while the band widths of the Br-4$p$ states in $Cmcm$- NaBr and NaBr[He] are 10.1 and 9.1 eV (Figure S12c). The increase in the volumes of the Na-Cl and Na-Br unit cells narrow the $p$-bands. He insertion into NaCl[He] and NaBr[He] lattices further narrows both the Cl-3$p$ and the Br-4$p$ bands significantly: their bands widths are 6.1 and 8.0 eV, respectively, in NaXHe. Thus, in Na(Cl,Br)He, the p states are much more localized following He insertion. Finally, Figure S12d shows that the occupied 5$p$ band widths of $Cmcm$-NaI and NaI[He] are 11.1 and 11.6 eV, respectively. Note that although these band widths are similar, the I-5$p$ band in NaI[He] extends above the Fermi level, as NaI[He] shows metallic character. He insertion in NaIHe re-introduces an electronic gap, with the I-5$p$ states fully occupied and the band width decreased to 10.6 eV. Thus, He insertion affects the valence $p$ states of Cl, Br and I significantly.

**Supplementary Section S8**: empirical comparison of interstitial site in anti-NiAs NaX lattice

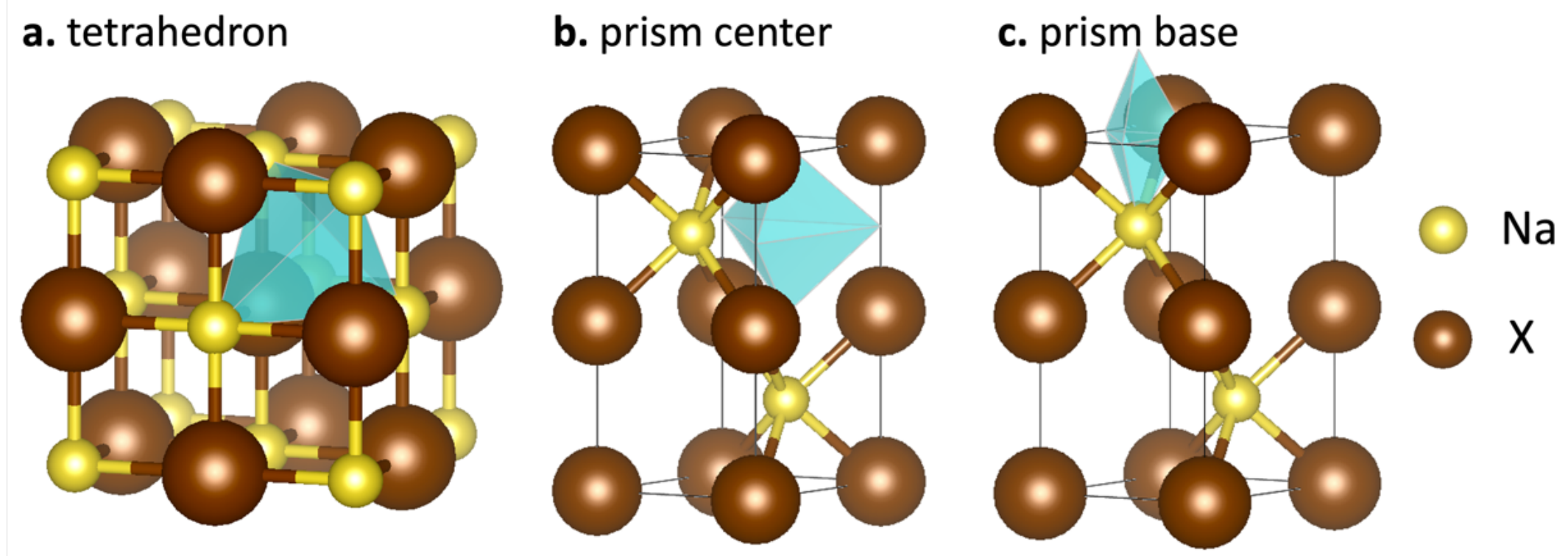

**Figure S13.** The interstitial site in *Fm*-3*m*-NaX and *P*6$_3$/*mmc*-NaX[He] in shape of certain polygons. **a**. the tetrahedral empty site in Fm-3m NaX structure. **b**. the prism center and c. prism base interstitial in *P*6$_3$/*mmc*-NaX[He].

The size of the interstitial sites shown in Figure S13 are measured by the volume of the polygons empirically. In Figure S13a, the interstitial site is in the shape of a tetrahedron. If the lattice constant of $Fm\bar{3}m$-NaX is set to be $a$, the volume of the tetrahedral interstitial then is $V_{\mathrm{tetra}} = \frac{a^3}{24}$. If *P*6$_3$/*mmc*-NaX[He] lattice constants are set to be $a = b$ and $c$, the prism centered (pc) interstitial site volume is $V_{\mathrm{pc}} = \frac{\sqrt{3}a^2c}{24}$ and the prism based (pb) interstitial site volume is $V_{\mathrm{pb}} = \frac{\sqrt{3}a^2c}{96}$. The prism centered interstitial site are 4 times large than the prism based ones.

**Supplementary Section S9:** Chemical pressure (CP)

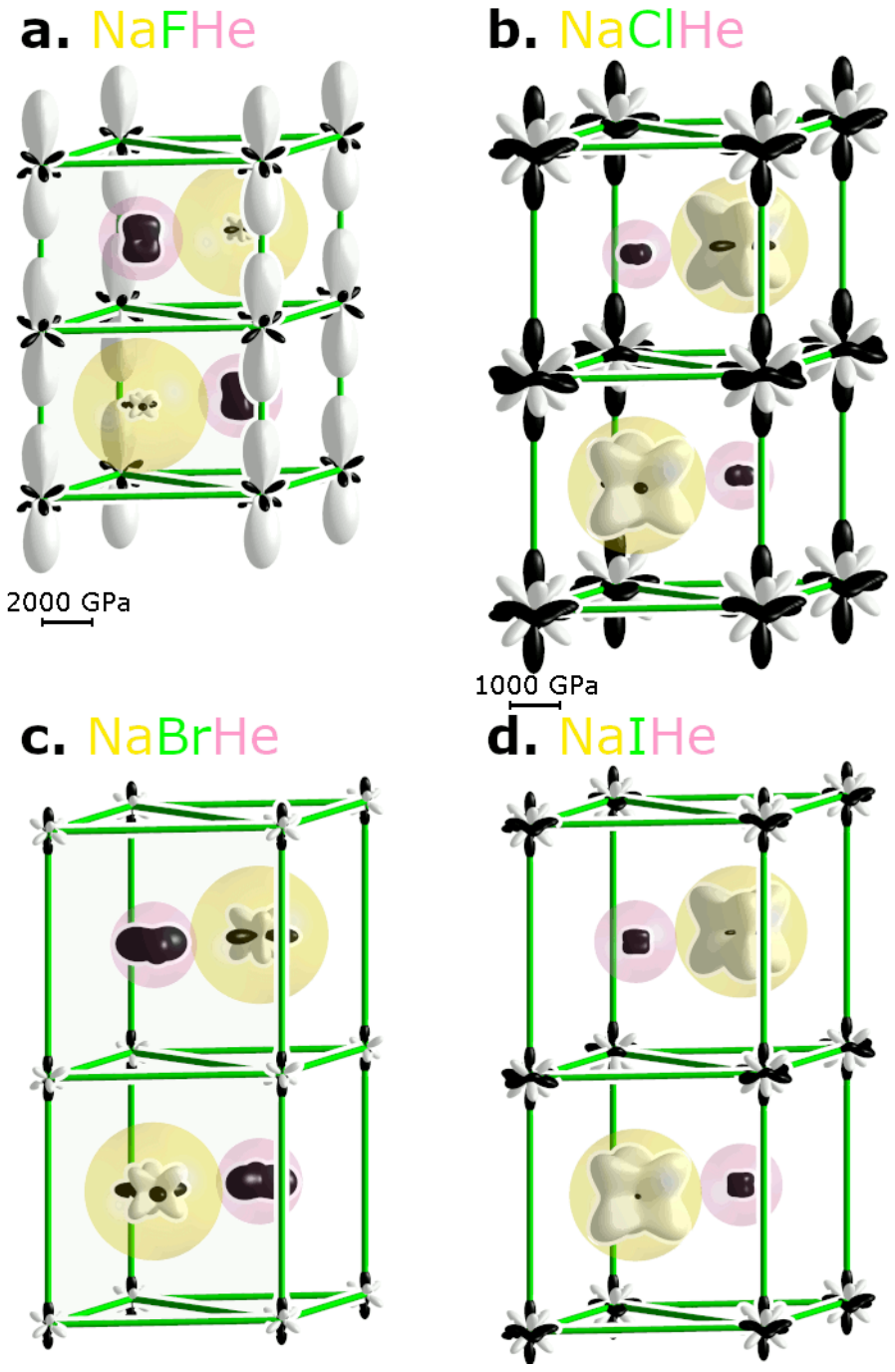

**Figure S14.** DFT-CP schemes for **a**. NaFHe, **b**. NaClHe, **c**. NaBrHe, and d. NaIHe at 300 GPa, all with $P6_3/mmc$ symmetry. For clarity, the CP scheme for NaFHe has been drawn at half scale relative to the others. Positive CP (denoting a desire for expansion of the coordination environment) is indicated by white lobes, while negative CP (denoting a desire for contraction) is indicated by black lobes.

We employed the DFT-Chemical Pressure (CP) analysis[4] to visualize the internal tensions driven by atomic packings in the various NaXHe phases. In a DFT-CP scheme, these tensions are represented by atom-centered spherical harmonics, where the magnitude of the CP in a particular direction is proportional to the size of the spherical harmonic lobe and the sign is reflected in the color of the lobe. Negative pressure, indicating a local desire for contraction, is represented by black lobes (in an analogy to black holes), while positive pressure, indicating a local desire for expansion, is represented by white lobes. Further details are provided in the supporting information. The DFT-CP method has recently been used successfully to understand the structural stability and distortions in high-pressure systems.[5,6]

In the CP schemes for the $P6_3/mmc$ NaXHe phases, the Na atoms (with the lobes inside yellow spheres drawn in Figure S14) are surrounded primarily by positive CP values, while the He atoms (lobes drawn inside pink spheres) are surrounded by negative CPs. Positive CPs stem from repulsive interactions of the ion core with the surroundings, and are lacking for the He atom in all cases. The general features of the CP surrounding both the Na and He atoms are similar for all of the phases considered. However, when considering the CP displayed by the X atoms, an immediate difference between the features present within NaFHe and the other phases containing heavier halides can be seen.

On the F atoms (Figure S14a), prominent positive CPs are directed towards the short F-F contacts along the *c* axis, while very small negative CP lobes are directed towards the nearby He atoms. This is interpreted as the fluoride ions being too close to each other along *c*. For the Cl, Br, and I atoms the trend is reversed (Figure S14b-d), with *negative* CP on the X-X contacts along the *c*-axis and in the *ab* plane, but positive CP towards the Na atoms, meaning that while these phases would prefer to contract along the X-X contacts, this contraction is stymied by the positive CP already present along the X-Na contact. The overly-compressed F-F contacts in $P6_3/mmc$ NaFHe instead highlight a cramped environment and an unsatisfactory structural arrangement.

## Supplementary Section S10: Energy decomposition analysis

**Table S8**. Kinetic, Coulomb and the exchange-correlation components of the energy terms comprising the Zielger-Rauk energy decomposition analysis,[7,8] calculated for the reaction NaXHe – (NaX[He] + [NaX]He) at 200 GPa, in [eV/f.u], at the BP86-D4/TZP level of theory.

| **BP86-D4/TZP** | **NaFHe** ***P6₃/mmc*** | **NaFHe** ***Pnma*** | **NaClHe** ***P6₃/mmc*** | **NaBrHe** ***P6₃/mmc*** | **NaIHe** ***P6₃/mmc*** |
|---|---|---|---|---|---|
| **Steric Interaction** | | | | | |
| Kinetic | 35.25 | 35.87 | 34.17 | 36.00 | 41.22 |
| Coulomb | -25.21 | -25.70 | -24.25 | -26.17 | -29.57 |
| Exchange-Correlation | -6.18 | -6.25 | -5.78 | -5.72 | -6.19 |
| $\Delta E_{Pauli} + \Delta E_{Elec}$ | 3.86 | 3.92 | 4.14 | 4.11 | 5.46 |
| **Orbital interaction** | | | | | |
| Kinetic | -18.62 | -18.95 | -18.69 | -20.97 | -26.43 |
| Coulomb | 15.73 | 16.03 | 15.64 | 17.66 | 20.93 |
| Exchange-Correlation | 2.26 | 2.28 | 2.14 | 2.26 | 2.87 |
| $\Delta E_{Orb}$ | -0.63 | -0.65 | -0.93 | -1.05 | -2.63 |

**Supplementary Section S11**: Ewald sum rule implemented in a python code

To compute the Madelung energy of the lattice of NaX and NaXHe, we developed a code by employing Ewald sum rule. We compare the Madelung energy between result from VESTA and from our code. In addition, we also compare the Ewald energy from VASP package and our code. The results are in Table S9.

It is worth to emphasize the difference between Madelung energy and Ewald energy. Madelung energy describes the interaction of all the anion and cations in the lattice. The electrons and nucleus are reduced into point charges. However, the Ewald energy refers the interaction of positive charged ion cores in the background of electron gas. To test the validity of our code, we compare the Madelung energy and Ewald energy of our codes with VESTA[9] and VASP package, respectively. We use NaCl of rock salt structure as example. This structure was in conventional cell whose formula is $Na_4Cl_4$, and was relaxed between 0 GPa and 300 GPa with a step of 50 GPa. For the computation of Madelung energy, Na and Cl are selected to be charged with +1 and -1. In Table S9, the results of VESTA and our code agree with each other. For Ewald energy, we use Na_sv and Cl_h POTCARs for the computation, which assume that the configuration of Na and Cl are $2s^2 2p^6 3s^1$ and $3s^2 3p^5$, respectively. Thus, Na and Cl are positively charged with +9 and +7. Our code also reproduces the Ewald energy value of VASP package, see also Table S9.

**Table S9**. The Madelung energy and Ewald energy of B1-NaCl between 0 GPa and 300 GPa.

| Pressure (GPa) | Lattice Constant (Å) | Madelung energy (eV/cell) | | Ewald Energy (eV/cell) | |
|---|---|---|---|---|---|
| | | VESTA | Our code | VASP | Our code |
| 0 | 5.668 | -35.520 | -35.517 | -3726.046 | -3726.010 |
| 50 | 4.717 | -42.680 | -42.676 | -4477.160 | -4477.118 |
| 100 | 4.442 | -45.328 | -45.325 | -4755.054 | -4755.009 |
| 150 | 4.274 | -47.106 | -47.102 | -4941.422 | -4941.375 |
| 200 | 4.153 | -48.473 | -48.469 | -5084.893 | -5084.844 |
| 250 | 4.059 | -49.597 | -49.594 | -5202.907 | -5202.858 |
| 300 | 3.981 | -50.563 | -50.562 | -5304.404 | -5304.354 |